\documentclass[12pt,preprint]{aastex}

\usepackage{url}
\usepackage[dvips]{color}
\usepackage[latin1]{inputenc}
\usepackage[english]{babel}
\usepackage{graphicx}
\usepackage{times}
\usepackage{amssymb}
\usepackage{color}
\usepackage[figuresright]{rotating}
\usepackage{amsmath}
\usepackage{longtable}
\usepackage{array}
\usepackage{ifthen}

\newcommand{\figurewidth} {0.45\textwidth}

\newcommand{\ha} {H$\alpha$}
\newcommand{\hII} {$\mathrm{H_{II}}$}
\newcommand{\Deg}{${}^{\circ}$}

\newcommand{\Sec}{${}^{\prime\prime}$}
\newcommand{\cic}{0.0023}
\newcommand{\cicAm}{0.0025}

\newcommand{\colorFigures}{1}

\shorttitle{EMCCD faint flux imaging}
\shortauthors{O. Daigle et al.}

\begin{document}


\title{Extreme faint flux imaging with an EMCCD}


\author{Olivier Daigle\altaffilmark{1,2,5}, Claude Carignan\altaffilmark{1,4}, Jean-Luc Gach\altaffilmark{2}, Christian Guillaume\altaffilmark{3}, Simon Lessard\altaffilmark{5}, Charles-Anthony Fortin\altaffilmark{5} and S\'ebastien Blais-Ouellette\altaffilmark{5}}
\email{odaigle@astro.umontreal.ca}

\altaffiltext{1}{Laboratoire d'Astrophysique Exp\'erimentale, D\'epartement de Physique, Universit\'e de Montr\'eal, C.P. 6128, succ. centre-ville, Montr\'eal, Québec, Canada, \mbox{H3T 2B1}}
\altaffiltext{2}{Aix-Marseille Universit\'e -- CNRS -- Laboratoire d'Astrophysique de Marseille, Observatoire Astronomique de Marseille-Provence, Technop\^ole de Ch\^ateau-Gombert,  38, rue Fr\'ed\'eric Joliot-Curie, 13388 Marseille, France}
\altaffiltext{3}{CNRS -- Observatoire Astronomique de Marseille-Provence -- Observatoire de Haute-Provence, 04870 St-Michel l'observatoire, France}
\altaffiltext{4}{Observatoire d'Astrophysique de l'Université de Ouagadougou, BP 7021, Ouagadougou 03, Burkina Faso}
\altaffiltext{5}{Photon etc., 5155 Decelles Avenue, Pavillon J.A Bombardier, Montr\'eal, Qu\'ebec, Canada, \mbox{H3T 2B1}}






\begin{abstract}
An EMCCD camera, designed from the ground up for extreme faint flux imaging, is presented. CCCP, the CCD Controller for Counting Photons, has been integrated with a CCD97 EMCCD from e2v technologies into a scientific camera at the Laboratoire d'Astrophysique Exp\'erimentale (LAE), Universit\'e de Montr\'eal. This new camera achieves sub-electron read-out noise and very low Clock Induced Charge (CIC) levels, which are mandatory for extreme faint flux imaging. It has been characterized in laboratory and used on the Observatoire du Mont M\'egantic 1.6-m telescope. The performance of the camera is discussed and experimental data with the first scientific data are presented.

\end{abstract}





\section{Introduction}

The advent of Electron Multiplying Charge Coupled Devices (EMCCD) allows sub-electron read-out noise to be achieved. However, the multiplication process involved in rendering this low noise level is stochastic. The statistical behaviour of the gain that is generated by the electron multiplying register adds an excess noise factor (ENF) that reaches a value of $2^{1/2}$ at high gains \citep{stanford}. The effect on the signal-to-noise ratio (SNR) of the system is the same as if the quantum efficiency (QE) of the EMCCD were halved. In this regime, the EMCCD is said to be in Analog Mode (AM) operation.

Some authors proposed offline data processing to lower the impact of the ENF \citep{2008MNRAS.386.2262L, 2003MNRAS.345..985B} in AM operation. However, one can overcome the ENF completely, without making any assumption on the signal's stability across multiple images, only by considering the pixel binary and by applying a single threshold to the pixel value. The pixel will be considered as having detected a single photon if its value is higher than the threshold and none if it is lower. In this way, the SNR will not be affected by the ENF and the full QE of the EMCCD can be recovered. In this regime, where the EMCCD is said to be in Photon Counting (PC) operation, the highest observable flux rate will be dictated by the rate at which the images are read-out; a frame rate that is too low will induce by losses by coincidence. 

However, at a high frame rate, the Clock Induced Charges (CIC) become dominant over the other sources of noise affecting the EMCCD (mainly dark noise). CIC levels in the range of 0.01 -- 0.1 were typically measured \citep{2008AIPC..984..148T, 2006SPIE.6276E..44W, techreport-minimal} on a 512$\times$512 CCD97 frame transfer EMCCD from e2v Technologies. Even at a low read-out speed of 1 frame s$\mathrm{^{-1}}$, these CIC levels are at least an order of magnitude higher than the dark noise. Thus, one wanting to do faint flux imaging with an EMCCD is stuck with two conflicting problems: a low frame rate is needed to lower the impact of the CIC whilst a high frame rate is needed if a reasonable dynamic range is to be achieved. 

In order to make faint flux imaging efficient with an EMCCD, the CIC must be reduced to a minimum. Some techniques have been proposed to reduce the CIC \citep{2006sda..conf..303T, 2004SPIE.5499..219D, 2004SPIE.5499..203M, 2004ASSL..300..611G, techreport-minimal, 2001sccd.book.....J} but until now, no commercially available CCD controller nor commercial cameras were able to implement all of them and get satisfying results. CCCP, the CCD Controller for Counting Photons, has been designed with the aim of reducing the CIC generated when an EMCCD is read out. It is optimized for driving EMCCDs at high speed ($\ge$ 10MHz), but may be used also for driving conventional CCDs (or the conventional output of an EMCCD) at high, moderate, or low speed. This new controller provides an arbitrary clock generator, yielding a timing resolution of $\sim$20ps and a voltage resolution of $\sim$2mV of the overlap of the clocks used to drive the EMCCD. The frequency components of the clocks can be precisely controlled, and the inter-clock capacitance effect of the CCD can be nulled to avoid overshoot and undershoots. Using this controller, CIC levels as low as 0.001 -- 0.002 event pixel$\mathrm{^{-1}}$ per frame were measured on the 512$\times$512 CCD97 operating in inverted mode. A CCD97 driven by CCCP was placed at the focus of the FaNTOmM instrument \citep{2002PASP..114.1043G, 2003SPIE.4841.1472H} to replace its GaAs photocathode-based Image Photon Counting System (IPCS). In this article, the important aspects of PC and AM operations with an EMCCD under low fluxes are outlined in section \ref{sect::faintFluxImaging}. In section \ref{sect::cccpPerf}, CCCP performance regarding these aspects is presented. Finally, in section \ref{sect::scientificResults}, scientific results obtained at the telescope are presented.

\section{Faint flux imaging with an EMCCD}
\label{sect::faintFluxImaging}

\subsection{The cost of sub-electron read-out noise}
The multiplication process that allows an EMCCD to reach sub-electron read-out noise is unfortunately stochastic. Only the \textit{mean} gain is known and it is not possible to know the \textit{exact} gain that was applied to a pixel's charge. This uncertainty on the gain and thus on the determination of the quantity of electrons that were accumulated in the pixel causes errors on the photometric measurements. This uncertainty can be translated in a SNR equation as an ENF, $F$, as follows:
\begin{equation}
\label{eqn::noiseFactor}
SNR = \dfrac{S}{\sqrt{F^2 S + \dfrac{\sigma_{real}^2}{G^2}}}.
\end{equation}
Here, $S$ is the quantity of electrons that were acquired and $G$ the mean gain achieved by the multiplication register. At high gain, $F^2$ reaches a value of 2 and its effect on the SNR is the same as if the QE of the EMCCD was halved \citep{stanford}.

\begin{figure*}[tbp]
\begin{center}
\includegraphics[width=\figurewidth]{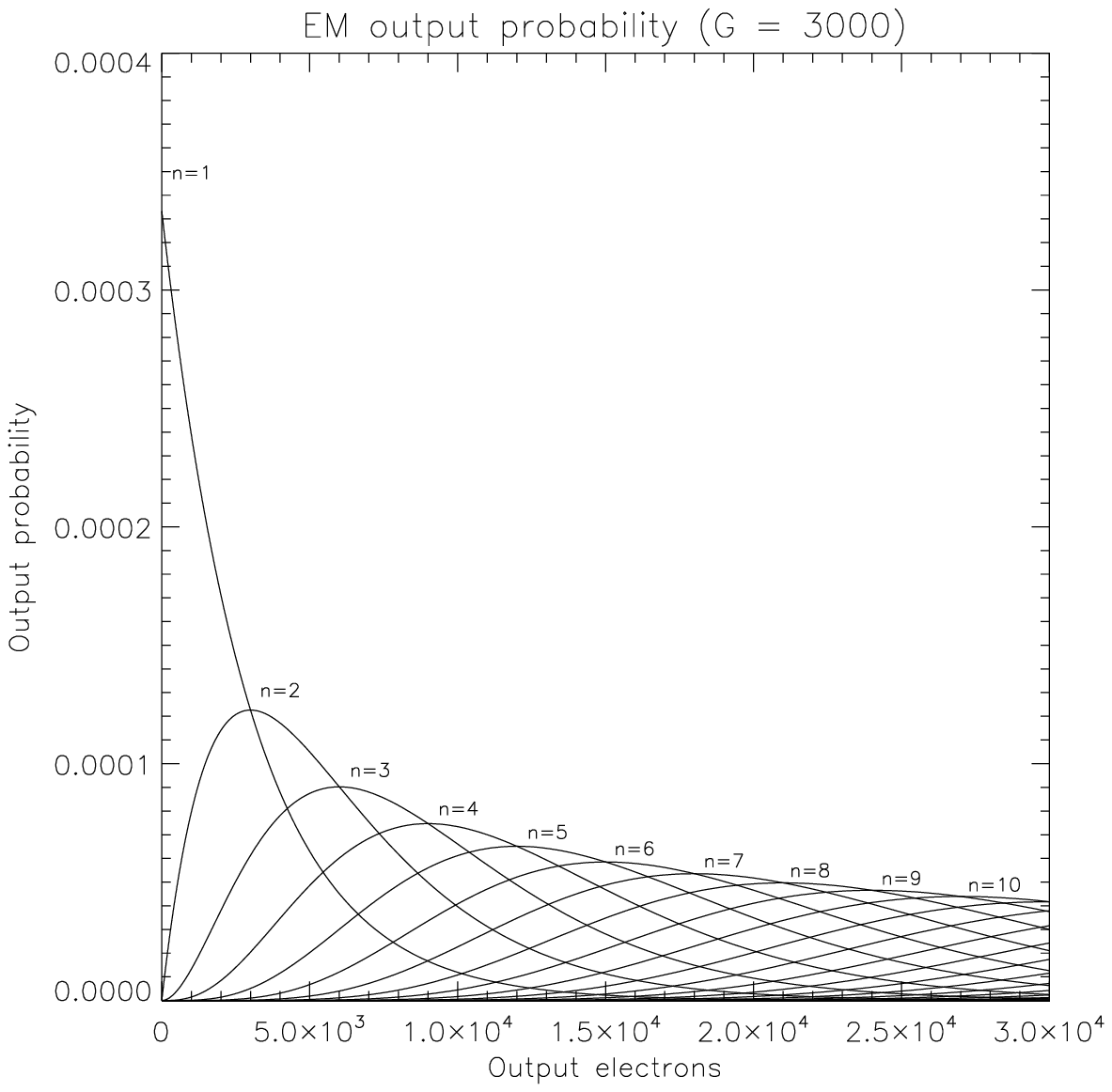}
\includegraphics[width=\figurewidth]{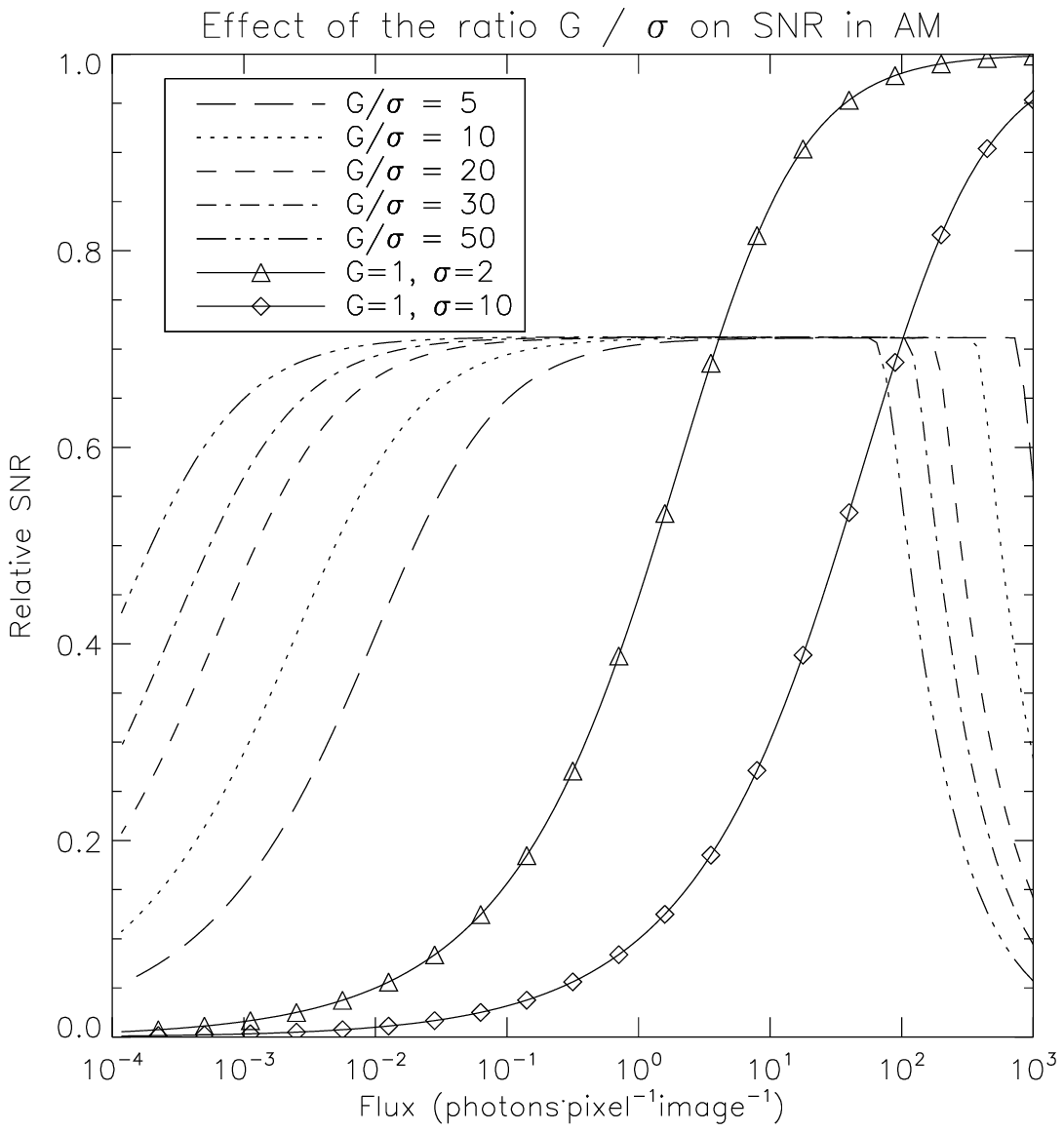}
\caption{\textbf{Left}: Output probability of the amplification register of an EMCCD, for a mean gain of 3000 plotted for various amount of input electrons, $n$. The overlapping regions are the result of the ENF induced by the multiplication process. \textbf{Right}: Relative SNR as a function of the photon flux per image, the read-out noise of the EMCCD and its operating gain. A saturation level of $200 000$ electrons is assumed, the read-out noise is 50 electrons unless noted, the dark noise and CIC are neglected.}
\label{fig::outputProb}
\end{center}
\end{figure*}

The effect of the ENF on the output probability of the EM stage of the EMCCD is outlined by figure \ref{fig::outputProb}, left panel. The overlapping output probabilities of different input electrons is the result of the ENF. Its impact on the SNR is shown in the right panel. The plateau at 0.707 of relative SNR is the effect of an ENF of value $\sqrt{2}$. This figure also compares the relative SNR of a conventional CCD ($G = 1$) operated at 2 and 10 electrons of read-out noise. 

Thus, for fluxes lower than $\sim$5 photons pixel$\mathrm{^{-1}}$ per image, the EMCCD in AM operation out-performs the low-noise ($\sigma$ = 2 electrons) conventional CCD. This does not take into account the duty cycle loss induced by the low speed readout of the low-noise CCD. Since most of the EMCCDs actually available are of frame transfer type, virtually no integration time is lost due to the read-out process and advantage would advantage would be even greater for the EMCCD compared to the conventional CCD. In AM operation, the maximum observable flux per image is a function of the EM gain of the EMCCD and the integration time should be chosen to avoid saturation.

\begin{figure*}[tbp]
\begin{center}
\includegraphics[width=\figurewidth]{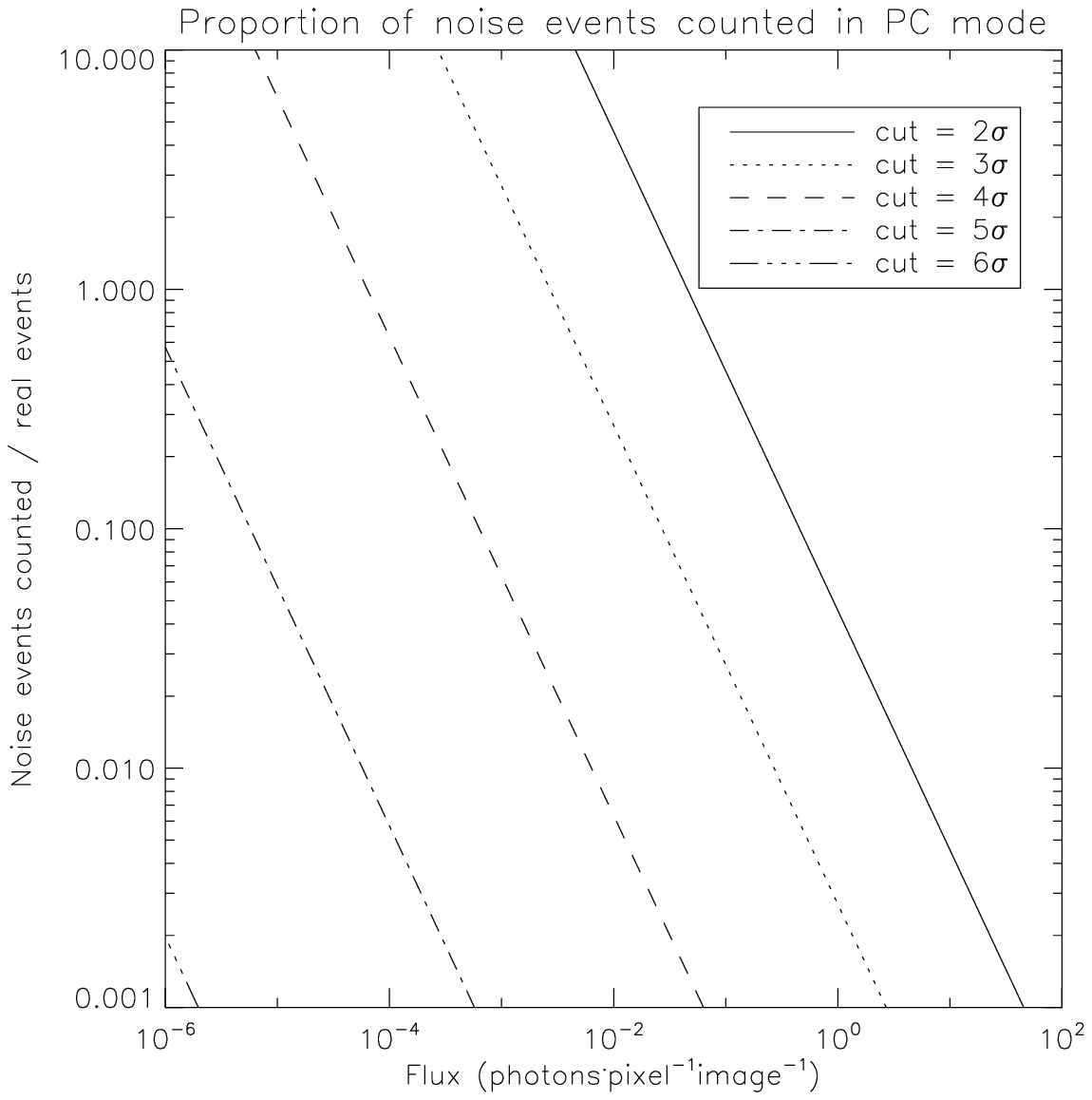}
\includegraphics[width=\figurewidth]{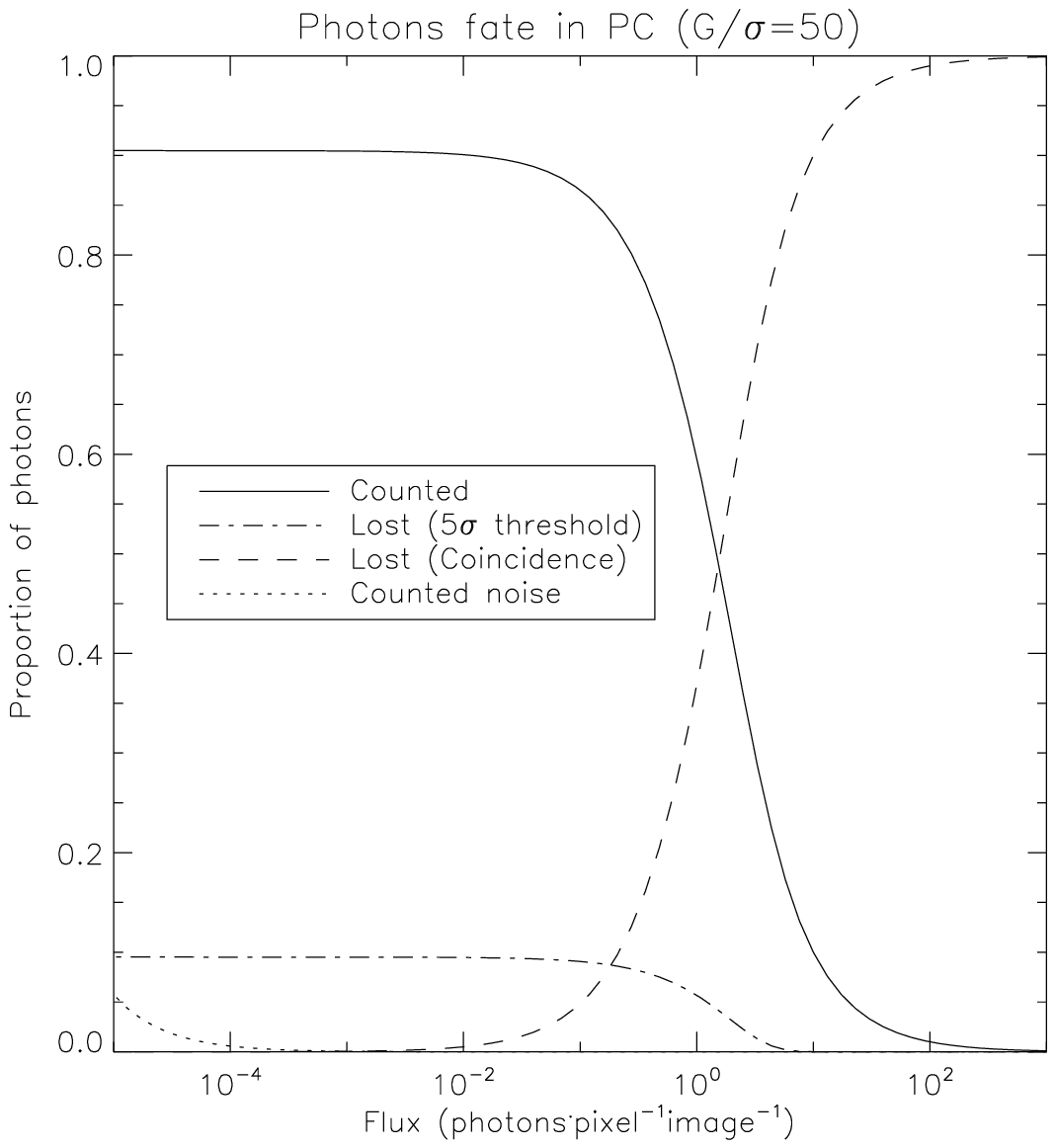}
\caption{\textbf{Left}: Proportion of noise events counted in photon counting mode as a function of the threshold, expressed in $\sigma$. \textbf{Right}: Photon fate in photon counting mode. The continuous line shows the proportion of counted photons, the dashed-dotted line shows the proportion of events lost in the read-out noise (due to the 5$\sigma$ threshold), the dashed line shows the proportion of events lost due to coincidence and the dotted line shows the proportion of false events that would be attributable to the read-out noise.}
\label{fig::photonFate}
\end{center}
\end{figure*}

This ENF affects the SNR only when one wants to measure more than one photon pixel$\mathrm{^{-1}}$. If one assumes that no more than one photon is to be accumulated, it can consider the pixel as being empty if the output value is lower than a given threshold, or filled by one photon if the output value is higher than the threshold. The threshold is determined solely by the real read-out noise of the EMCCD. Typically, a threshold of 5$\sigma$ allows one to avoid counting false events due to the read-out noise (see figure \ref{fig::photonFate}, left panel). Thus, even if the exact gain is still unknown, it has no impact since it does not enter in the equation of the output value. In this operating mode, the ENF, $F$, vanishes by taking a value of 1 \citep{stanford, 2006SPIE.6276E..42D}.

Counting only one photon pixel$\mathrm{^{-1}}$ does have its drawbacks: at high fluxes, coincidence losses become important (figure \ref{fig::photonFate}, right pannel, continuous and dashed line). The only way of overcoming the coincidence losses is to operate the EMCCD at a higher frame rate. In order to lose no more than 10\% of the photons by coincidence losses, the frame rate must be at least 5 times higher than the photon flux. This requires that the EMCCD be operated at high speed (typically $\ge$ 10MHz) in order to allow moderate fluxes to be observed.

\subsection{The need for high EM gain}
Figure \ref{fig::photonFate} gives a hint of another aspect of the PC operation: the ratio of the gain over the real read-out noise is important. Since PC operation involves applying a threshold that is chosen as a function of the real read-out noise, the EM gain that is applied to the pixel's charge will affect the amount of events that can be counted. Figure \ref{fig::outputProb} shows that for 1 input electron, the highest output probability of the EM register is 1 electron. This causes an appreciable proportion of events ending-up buried in the read-out noise. The proportion of events lost in the read-out noise, $e_l$, which is the proportion of the events that come out of the EMCCD at a value lower than $cut$ electrons, can be estimated by means of the following convolution:
\begin{equation}
\label{eqn::gainEffect}
e_l = \dfrac{\displaystyle\sum_{x=0}^{cut}{f(n,\lambda) \ast p(x,n,G)}}{1-f(0,\lambda)},
\end{equation}
where $f(n, \lambda)$ is the Poissonian probability of having $n$ photons during an integration period under a mean flux of $\lambda$ (in photon/pixel/frame) and $p(x,n,G)$ is the probability of having $x$ output electrons when $n$ electrons are present at the input of the EM stage at a gain of $G$. This probability is defined by
\begin{equation}
\label{eqn::emOutputProb}
p(x,n,G) = \dfrac{x^{n-1}e^{-x/G}}{G^{n}(n-1)!}.
\end{equation}

\begin{figure}[tb]
\begin{center}
\includegraphics[width=\figurewidth]{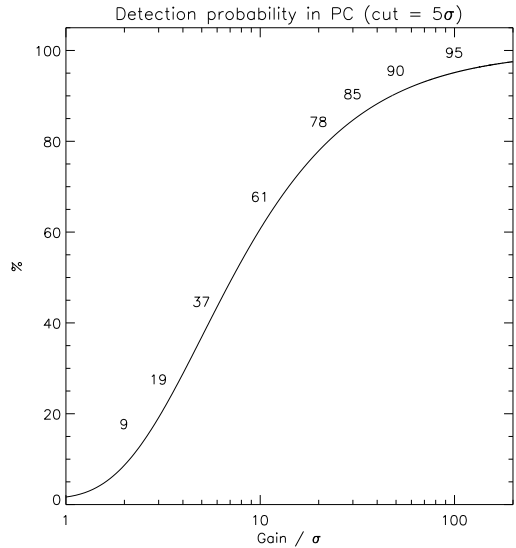}
\caption{Maximum proportion of counted photons as a function of the gain over read-out noise ratio, for a 5$\sigma$ threshold}
\label{fig::gainEffect}
\end{center}
\end{figure}

Figure \ref{fig::gainEffect} translates the importance of having a high gain over read-out noise ratio into a detection probability as a function of the $G/\sigma$ ratio. An EMCCD that is operated at an EM gain of 1000 and has a read-out noise of 50 electrons will allow no more than 78\% of the photons to be counted. In order to count 90\% of the photons, one would have to use an EM gain of 2500 for the same read-out noise. It is the ratio of the gain over the real read-out noise that sets the maximum amount of detected photons. Thus, even if an EMCCD is capable of sub-electron read-out noise, its effect is not completely suppressed. The read-out noise still affect the quality of the data if the EM gain is not high enough.

\subsection{The dominance of CIC}
When an EMCCD is operated under low fluxes at high gain and high frame rate, a noise source that is usually buried into the read-out noise of a conventional CCD quickly arises: CIC \citep{techreport-minimal}. The CIC are charges that are generated as the photo-electrons are moved across the CCD to be read out. The CIC noise may appear as dark noise except that it has no time component. It is only when the CCD is read out that the CIC occur. Thus, the higher the frame rate, the higher the CIC.

\begin{figure*}[tbp]
\begin{center}
\includegraphics[width=\figurewidth]{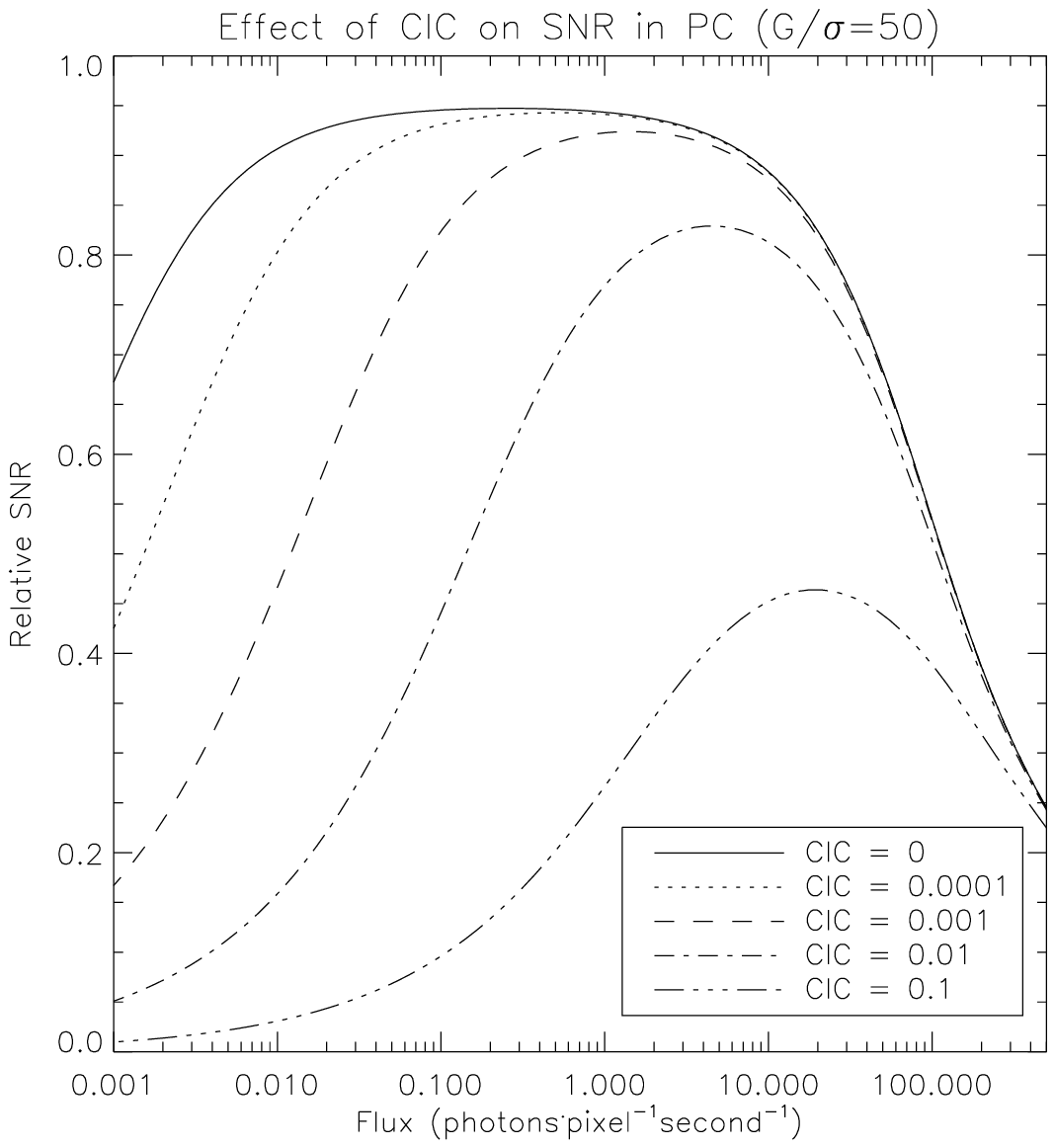}
\includegraphics[width=\figurewidth]{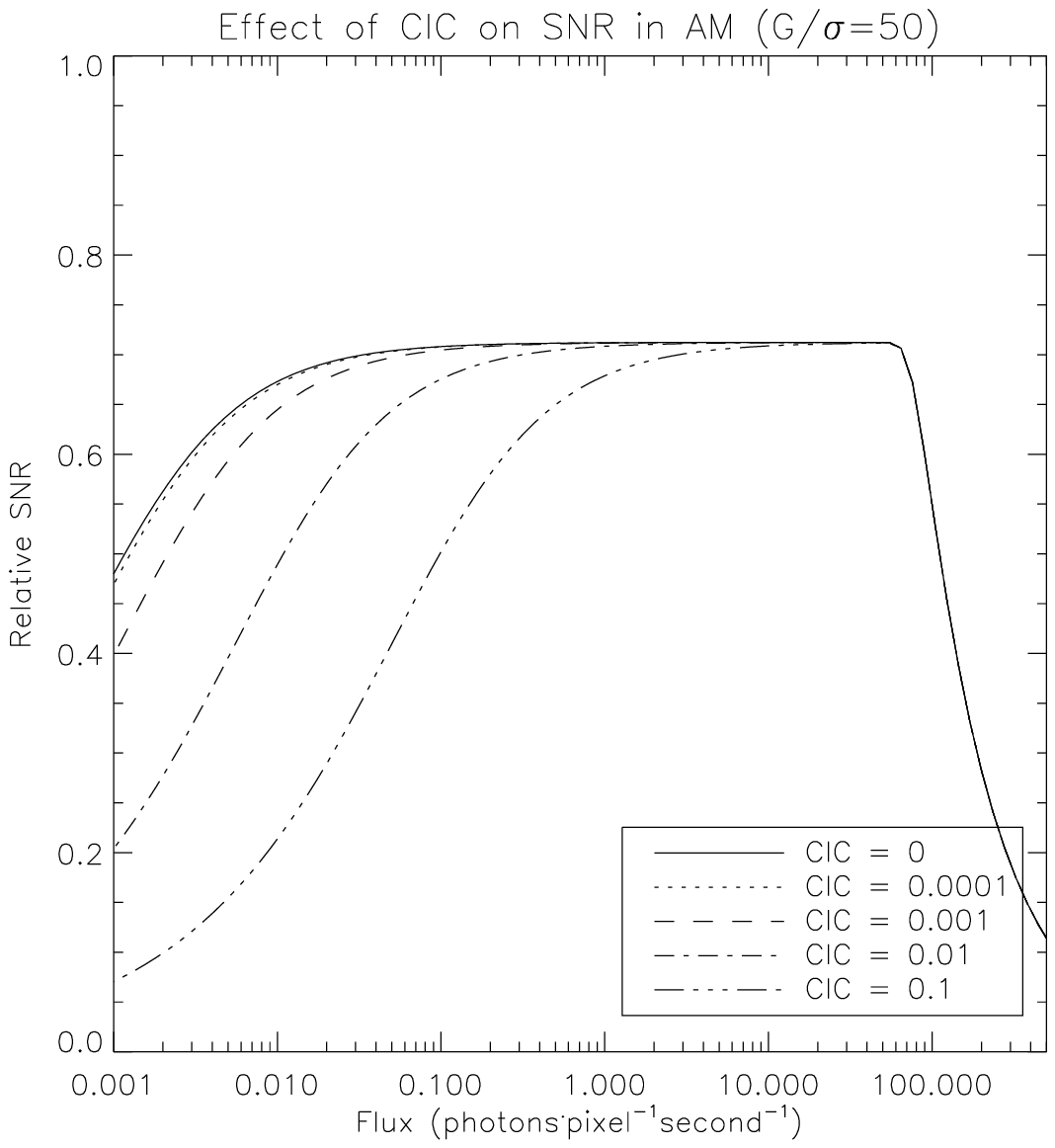}
\includegraphics[width=\figurewidth]{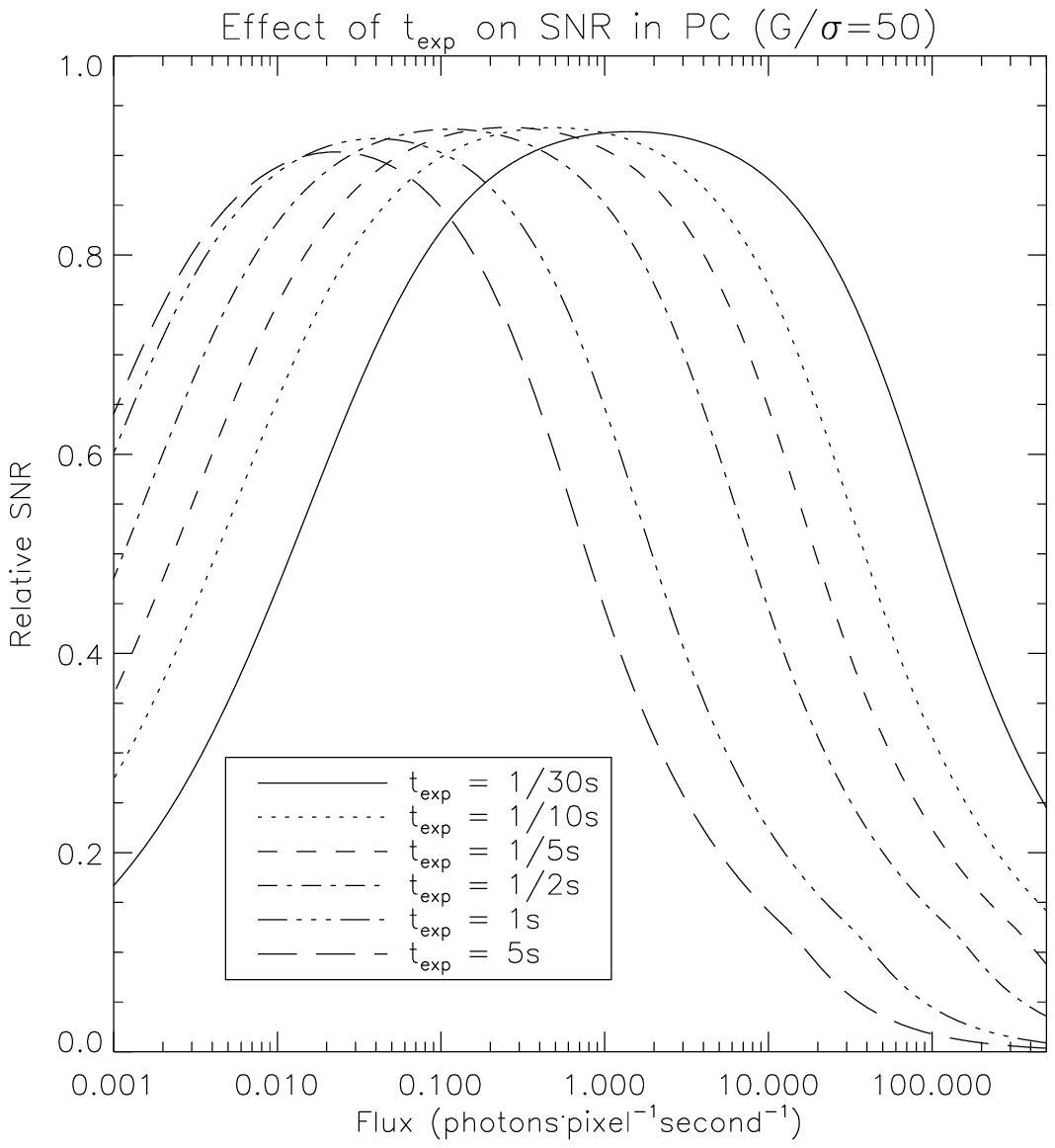}
\includegraphics[width=\figurewidth]{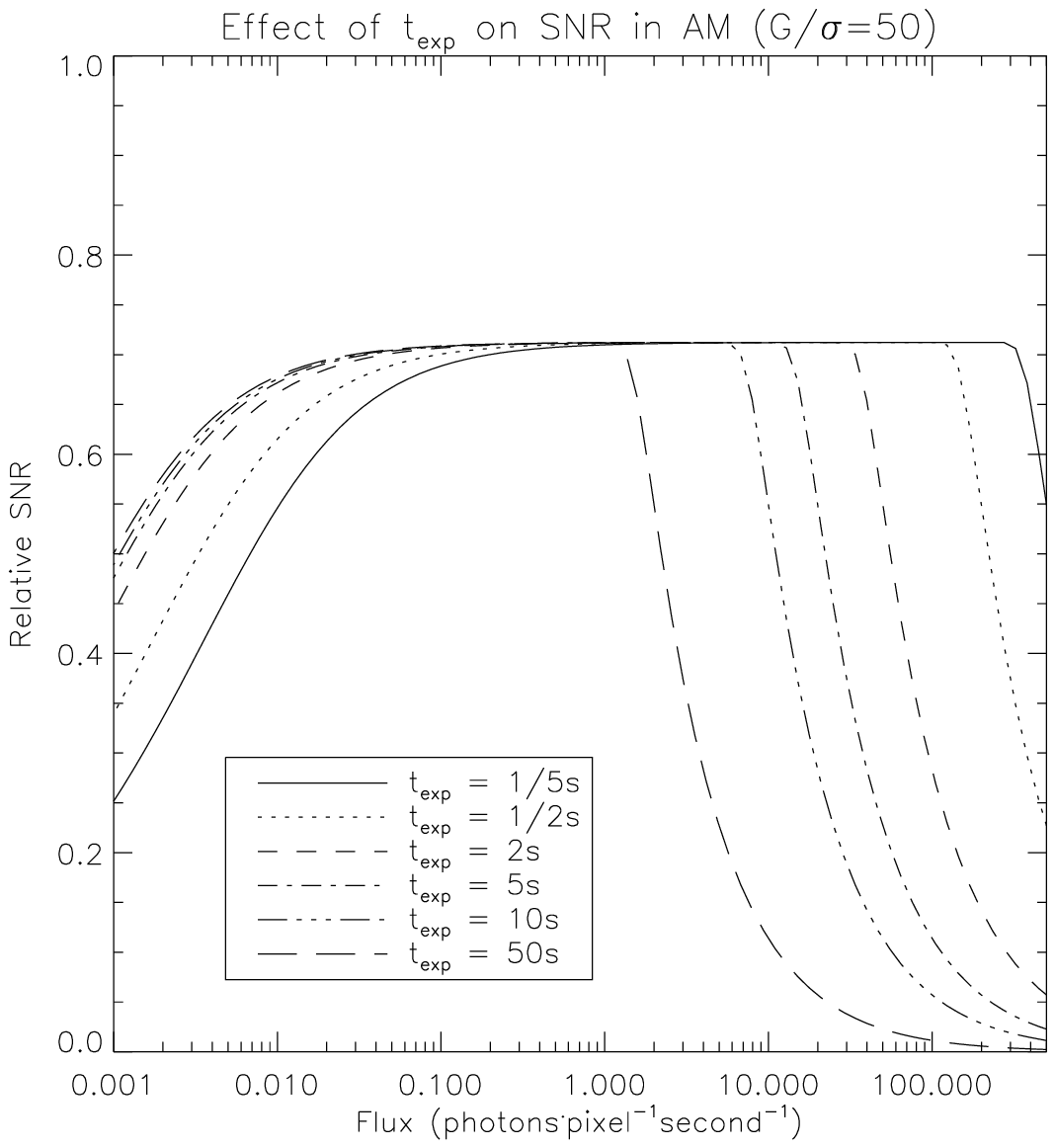}
\caption{Effect of various CIC levels (in electron per pixel per image) on the SNR. SNR is plotted against the one of a perfect photon counting device that would be affected only by shot noise. CIC rates are in electron per pixel per image, the dark noise is 0.001 electron per pixel per second, the EMCCDs are operated at a $G/\sigma$ ratio of 50 and a saturation level of 200000 electrons is assumed. \textbf{Top left}: EMCCD in PC operated at 30 frames per second with a threshold of 5$\sigma$. \textbf{Top right}: EMCCD in AM at 1 frame per second. \textbf{Bottom left}: EMCCD in PC, CIC is 0.001 event per pixel per frame and integration time is varying. \textbf{Bottom right}: EMCCD in AM, CIC is 0.001 event per pixel per frame and integration time is varying.}
\label{fig::cicComparison}
\end{center}
\end{figure*}

Figure \ref{fig::cicComparison} shows the effect of the CIC on the SNR of an EMCCD. One quickly realizes that, at 30 fps, the CIC dominate over the dark noise even for a CIC level as low as 0.0001 electron per pixel per image (top left panel). In AM, at 1 fps, the CIC dominate for levels higher than 0.001 (top right panel). In order to lower the impact of the CIC, one could chose to operate the EMCCD at a lower speed (higher integration time). The effect of this choice is outlined in the bottom panels of the figure (assuming a CIC level of 0.001). It shows that in PC and AM mode, there is little advantage of operating the EMCCD at less than $\sim$0.5 fps. For a given CIC level, the maximum exposure time should be dictated by the time needed for the dark noise to dominate (about twice as high as the CIC). Moreover, the losses due to saturation (in AM) or by coincidence (in PC) are becoming more important, and the rising of the integrating time reduces the dynamic range of the images without providing a gain in SNR at low flux. Thus, there is a minimum frame rate at which the EMCCD should be operated, either in AM or PC. For a lower frame rate, there is no further gain in SNR to achieve at low fluxes and losses occur at high fluxes.

Another problem with the CIC is that it is dependent of the EM gain (see section \ref{sect::cccpCic}). The higher the EM gain, the higher the rate at which the CIC is generated. Thus, in order to reach the high EM gain needed for faint flux imaging (30--50 times the read-out noise), one must really tame the CIC down to low levels.

\section{CCCP performance}
\label{sect::cccpPerf}

The camera built using CCCP and a CCD97 (hereafter CCCP/CCD97) was used to gather experimental data and to compare the performance of the CCCP controller against an existing, commercial, CCD97 camera (namely the, Andor \mbox{iXon$^{\mathrm{EM}}$+ 897} BI camera, hereafter the Andor camera). The CCCP/CCD97 performance is also presented in absolute numbers. All the data presented, for both cameras, were gathered at a pixel rate of 10 MHz.

\subsection{Real and effective read-out noise}
The real read-out noise of CCCP/CCD97 was measured at 10MHz of pixel rate using the photon transfer curve method with a faint, constant illumination and a varying integration time. For this technique, the variance of the images is plotted against the mean signal (read \citeauthor{2001sccd.book.....J} \citeyear{2001sccd.book.....J} for a detailed explanation of the technique). This technique yields both the read-out noise and the reciprocal gain (electrons ADU$\mathrm{^{-1}}$) of the CCD.

\begin{figure*}[tbp]
\begin{center}
\includegraphics[width=\figurewidth]{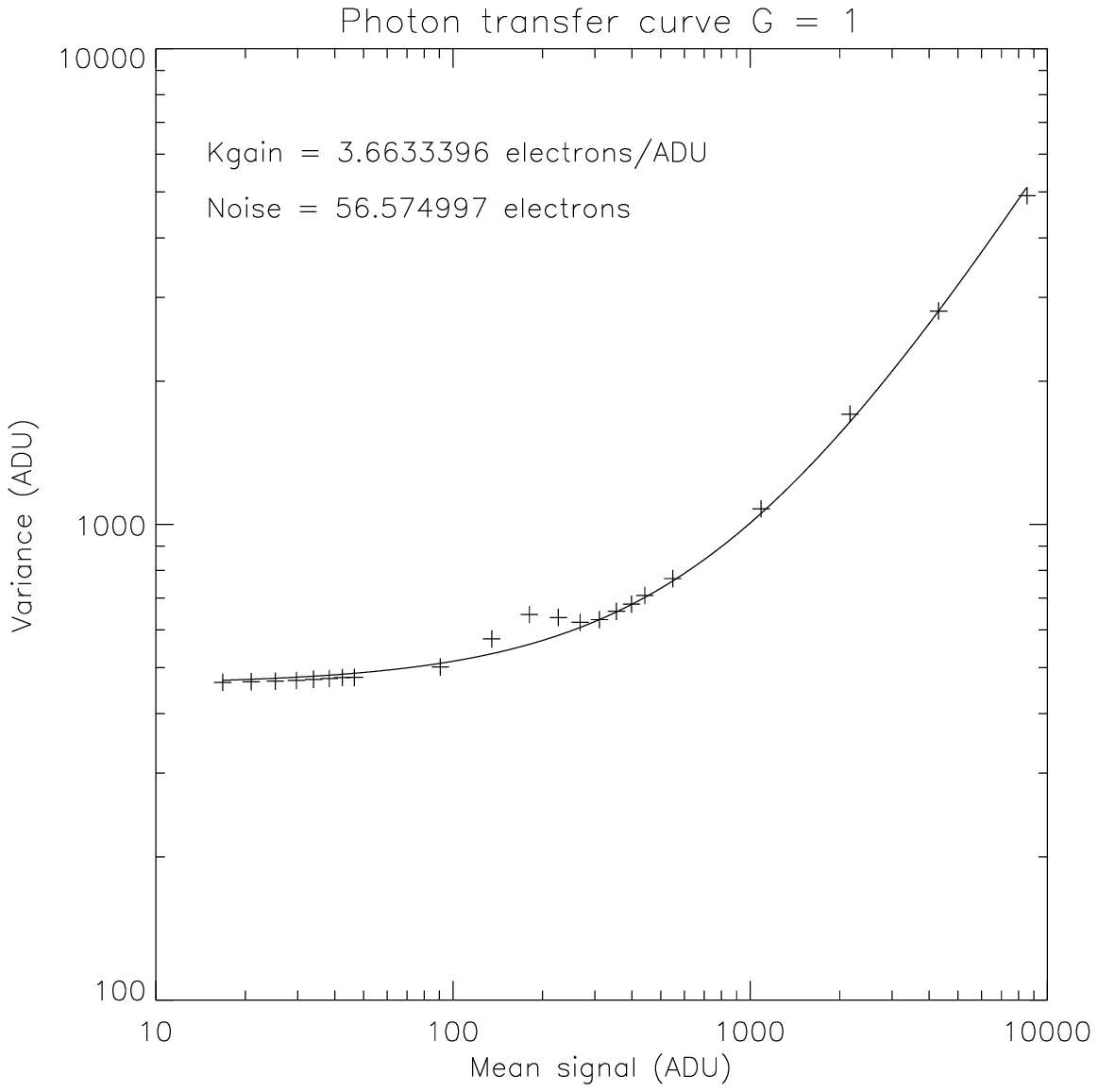}
\includegraphics[width=\figurewidth]{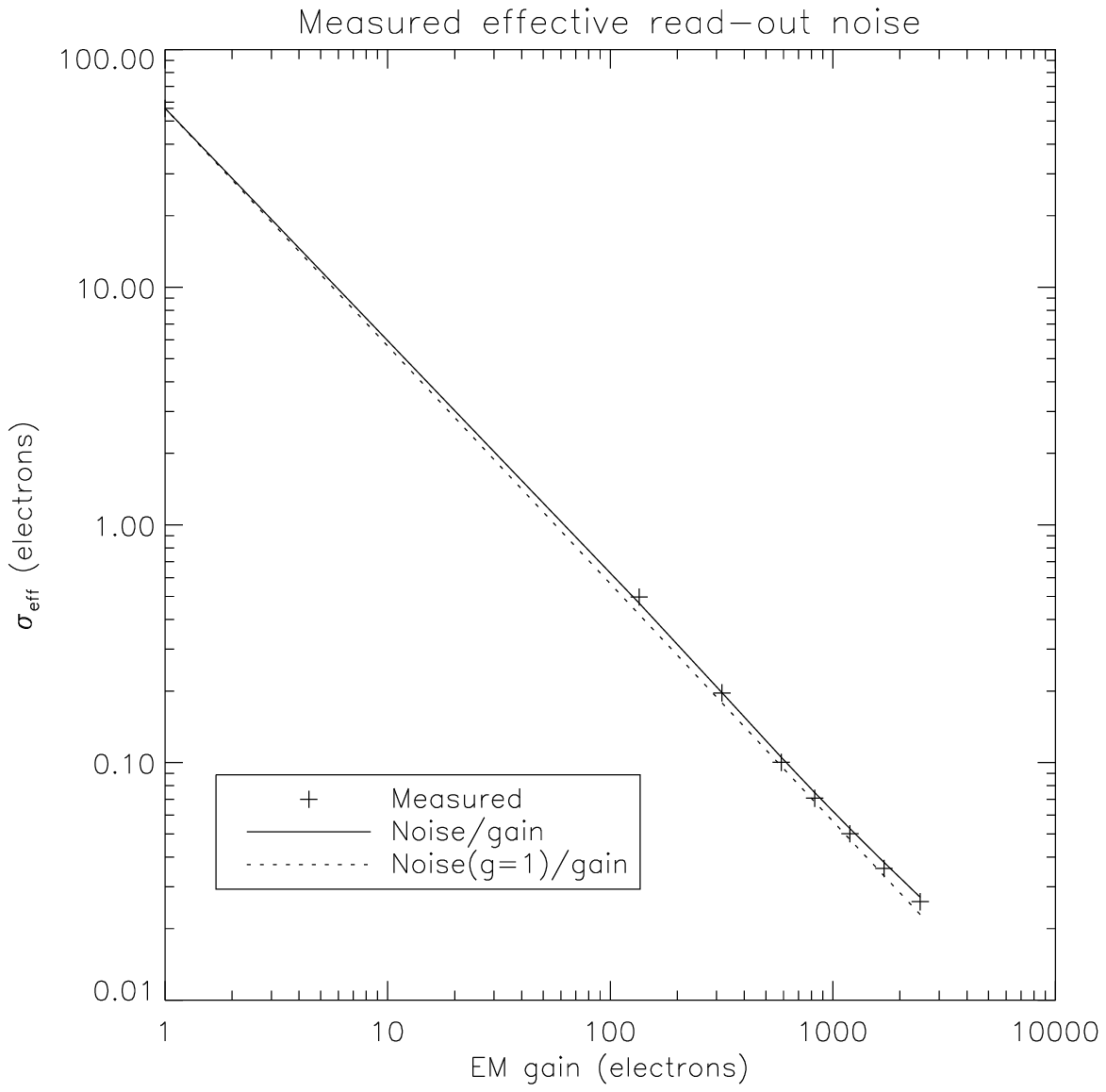}
\caption{Measurement of the real and effective read-out noise of CCCP/CCD97 at a pixel rate of 10MHz. \textbf{Left}: Photon transfer curve of CCCP/CCD97 at an EM gain of 1. The $+$ signs represent data and the curve represents the best fit to these data. \textbf{Right}: Effective read-out noise as a function of the EM gain. The $+$ signs represent the measured effective read-out noise, as measured by the photon transfer technique. The plain line represents the effective read-out noise expected by measuring the read-out noise (by taking the standard deviation of the signal) and dividing it by the measured EM gain. The dotted line represents the read-out noise measured at $G=1$ (left panel) and dividing it by the EM gain. See text for explanations.}
\label{fig::cccpRo}
\end{center}
\end{figure*}

Figure \ref{fig::cccpRo} shows the measurement for both the real and the effective read-out noise of CCCP /CCD97. The various plots of the left panel show that the real read-out noise is not dependent of the EM gain. Indeed, there is no reason why the read-out noise of the EMCCD could not be affected by the EM gain. The clocking of the High Voltage (HV) clock could induce substrate bounce, EMI or cross-talk and increase the real read-out noise. Moreover, the reciprocal gain of the charge domain (after amplification) of the EMCCD could also change if the output amplifier, for example, was not linear. The good concordance of the three data sets of the left panel shows that the real read-out noise and the reciprocal gain of the EMCCD are stable. One can assume that the effective read-out noise of the CCCP/CCD97 can be calculated by means of
\begin{equation}
\sigma_{eff} = \dfrac{\sigma_{G=1}}{G},
\end{equation}
where $\sigma_{G=1}$ is the read-out noise calculated at an EM gain of 1 and $G$ is the gain at which the EMCCD is operated.

\subsection{Clock Induced Charges}
\label{sect::cccpCic}

\begin{figure*}[tbp]
\begin{center}
\includegraphics[width=\figurewidth]{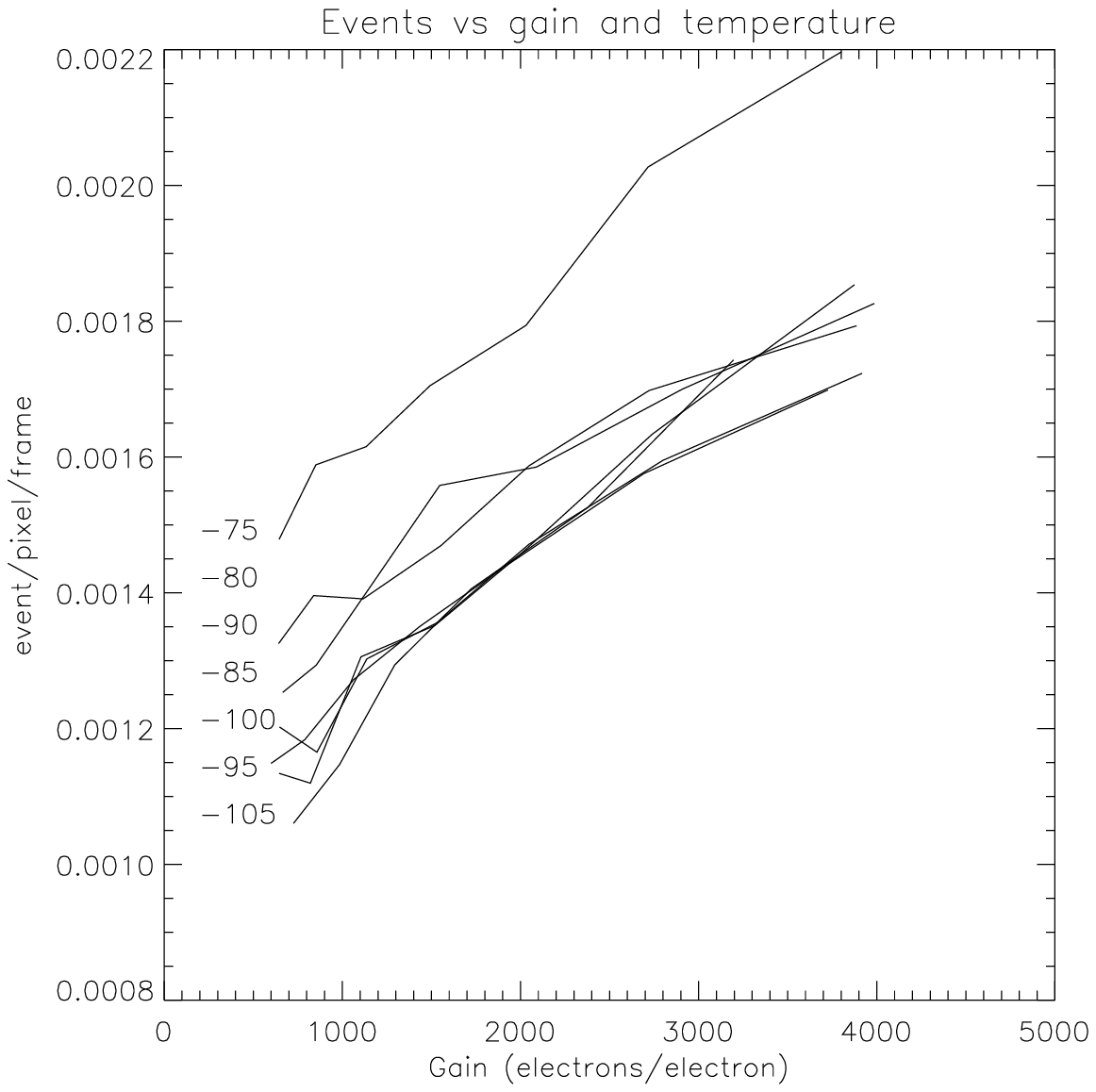}
\includegraphics[width=\figurewidth]{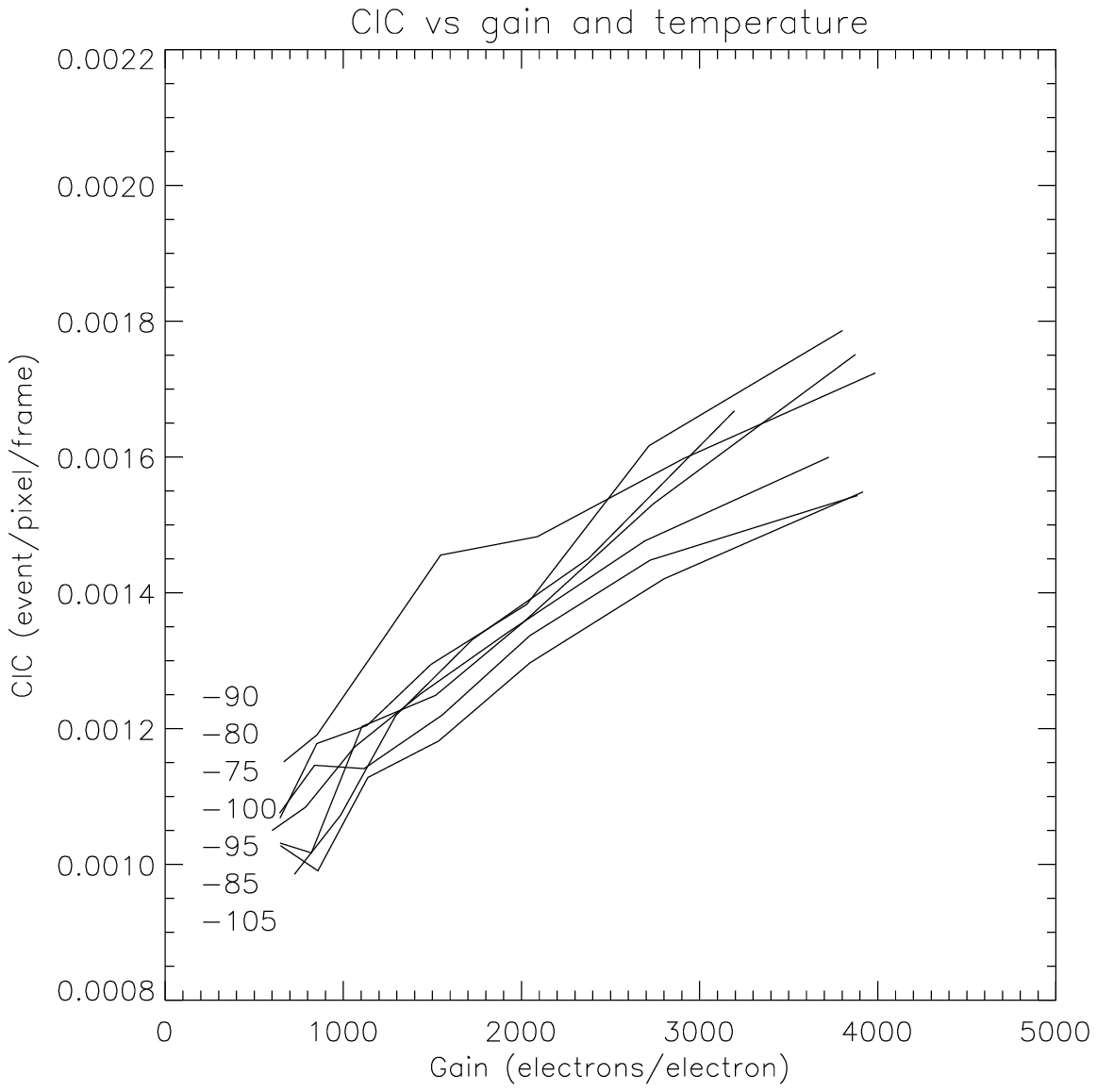}
\caption{Measurement of all the events generated during the read out process. \textbf{Left}: including dark noise. \textbf{Right}: excluding dark noise.}
\label{fig::cccpCic}
\end{center}
\end{figure*}

\begin{figure*}[tbp]
\begin{center}
\includegraphics[width=\figurewidth]{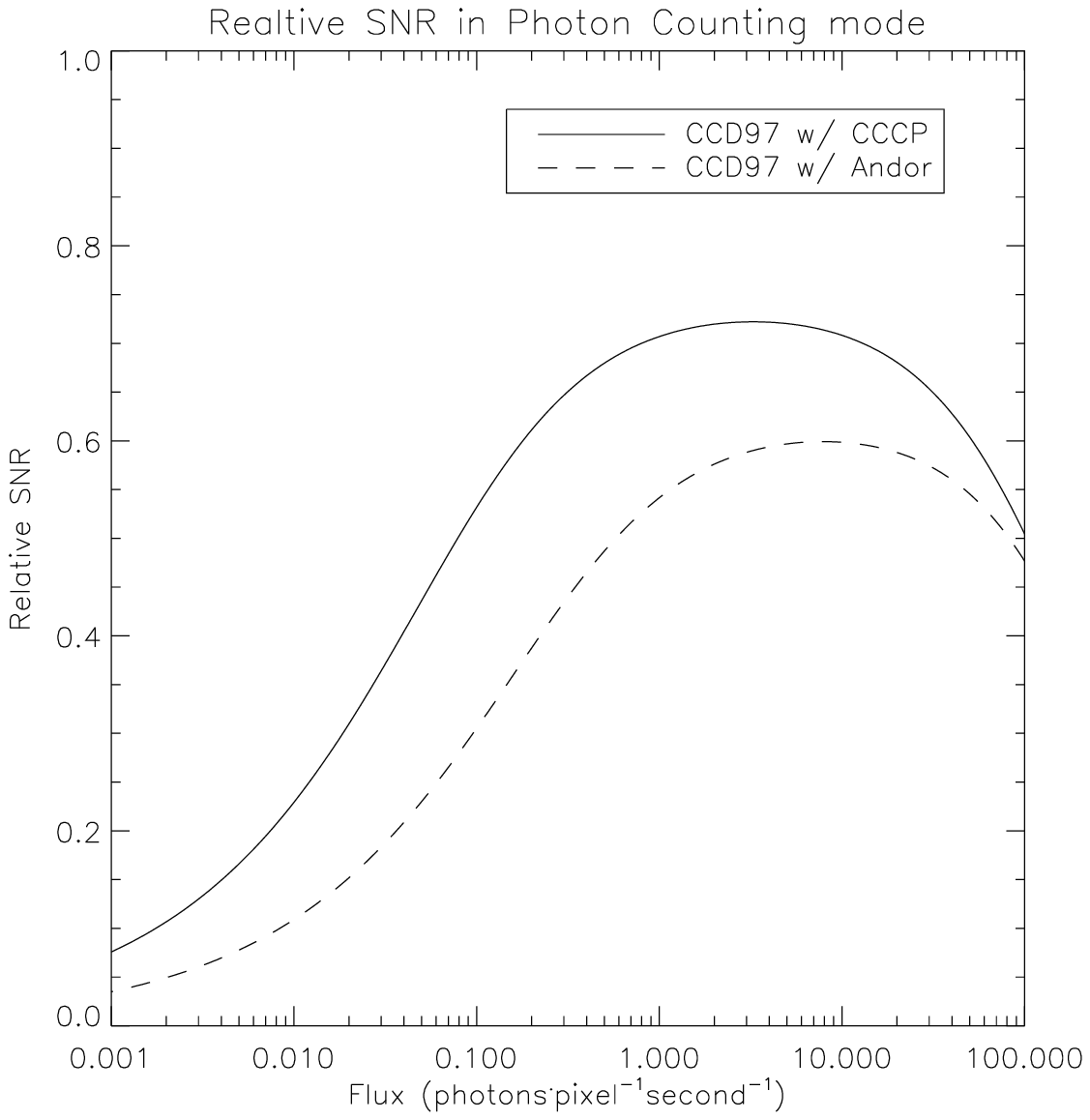}
\includegraphics[width=\figurewidth]{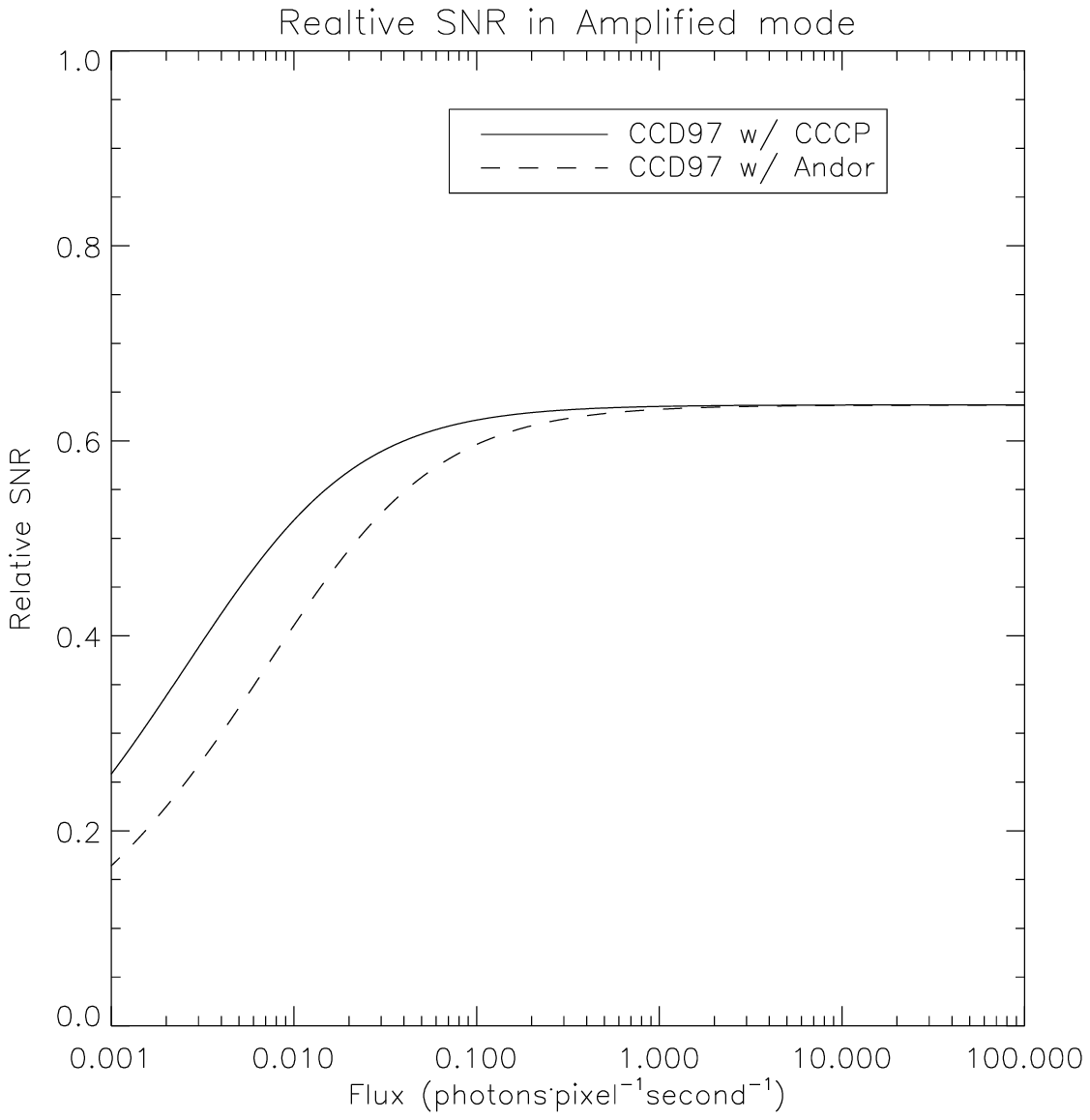}
\includegraphics[width=\figurewidth]{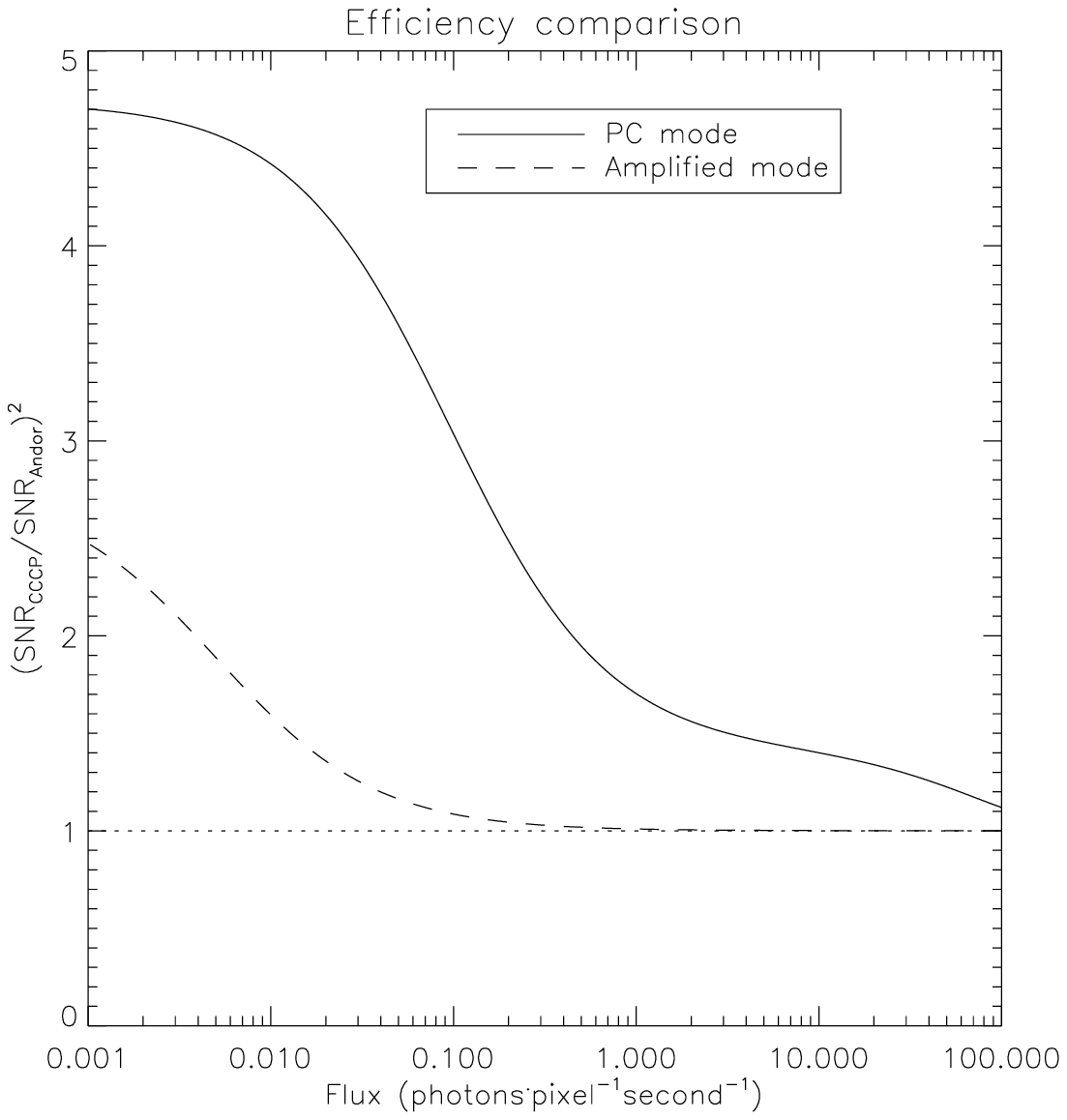}
\caption{Comparison of the SNR of a CCCP/CCD97 and the Andor camera. Comparisons are normalized with a perfect photon counting device. Both cameras have a QE of 80\%, a dark noise of 0.001 electron per pixel per second and a saturation level of 200000 electrons. For CCCP, the gain over read-out noise ratio is 30 and the CIC rate is \cic. For the commercial camera, the gain over read-out noise ratio is 16 and the CIC rate is 0.0084. In both cases, the effect of bad events is neglected. \textbf{Top left}: Cameras operated in PC. Both cameras are running at 30 frames per second. \textbf{Top right}: Cameras operated in AM. Both cameras are running at 1 frame per second. \textbf{Bottom}: Efficiency comparison, showing the ratio of the time needed to reach a given SNR at a given flux, between CCCP/CCD97 and the Andor Camera.}
\label{fig::snComparison}
\end{center}
\end{figure*}

\begin{figure*}[htbp]
\begin{center}
\ifthenelse{\colorFigures=1} {
\includegraphics[width=\figurewidth]{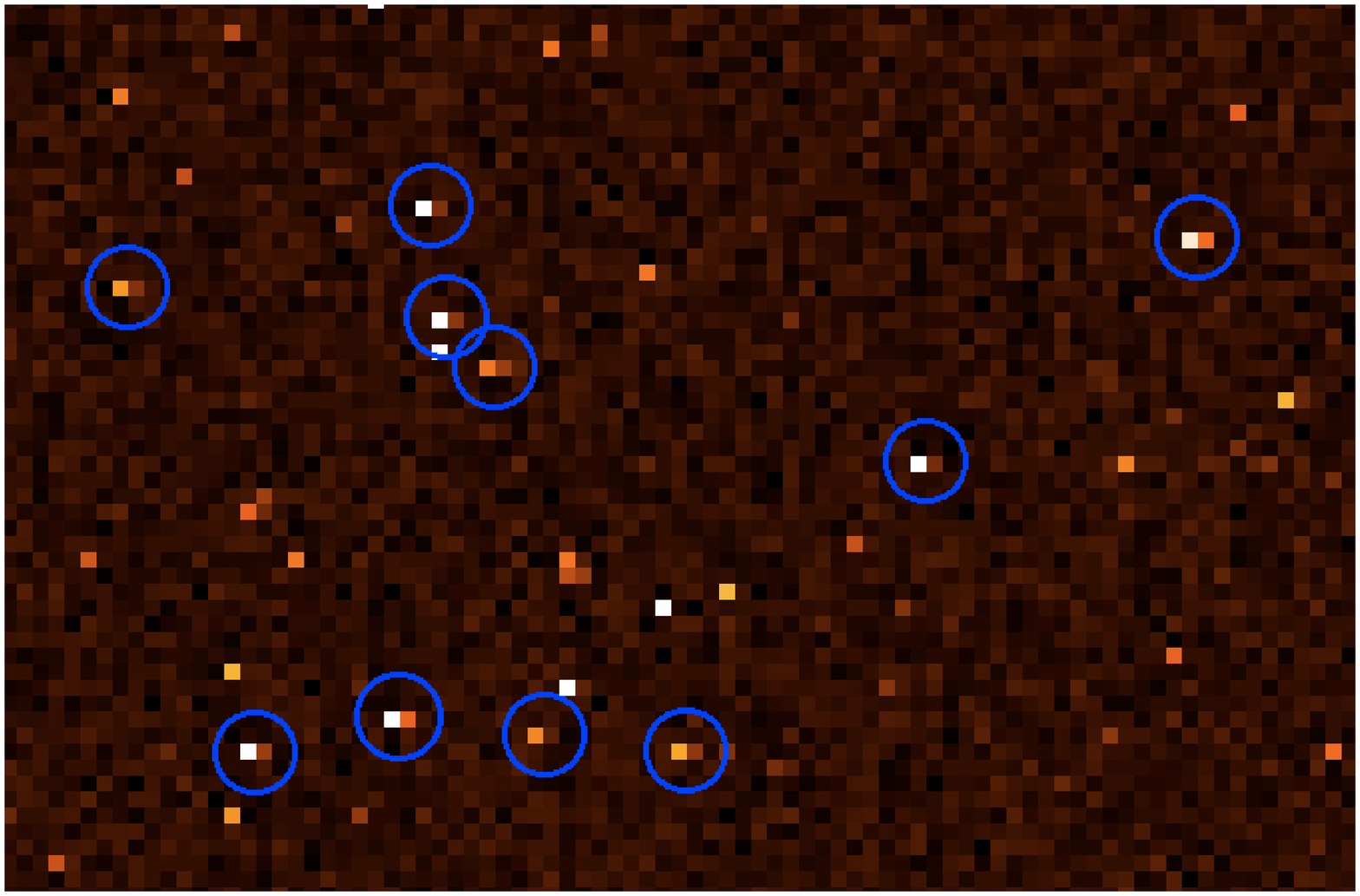}
\includegraphics[width=\figurewidth]{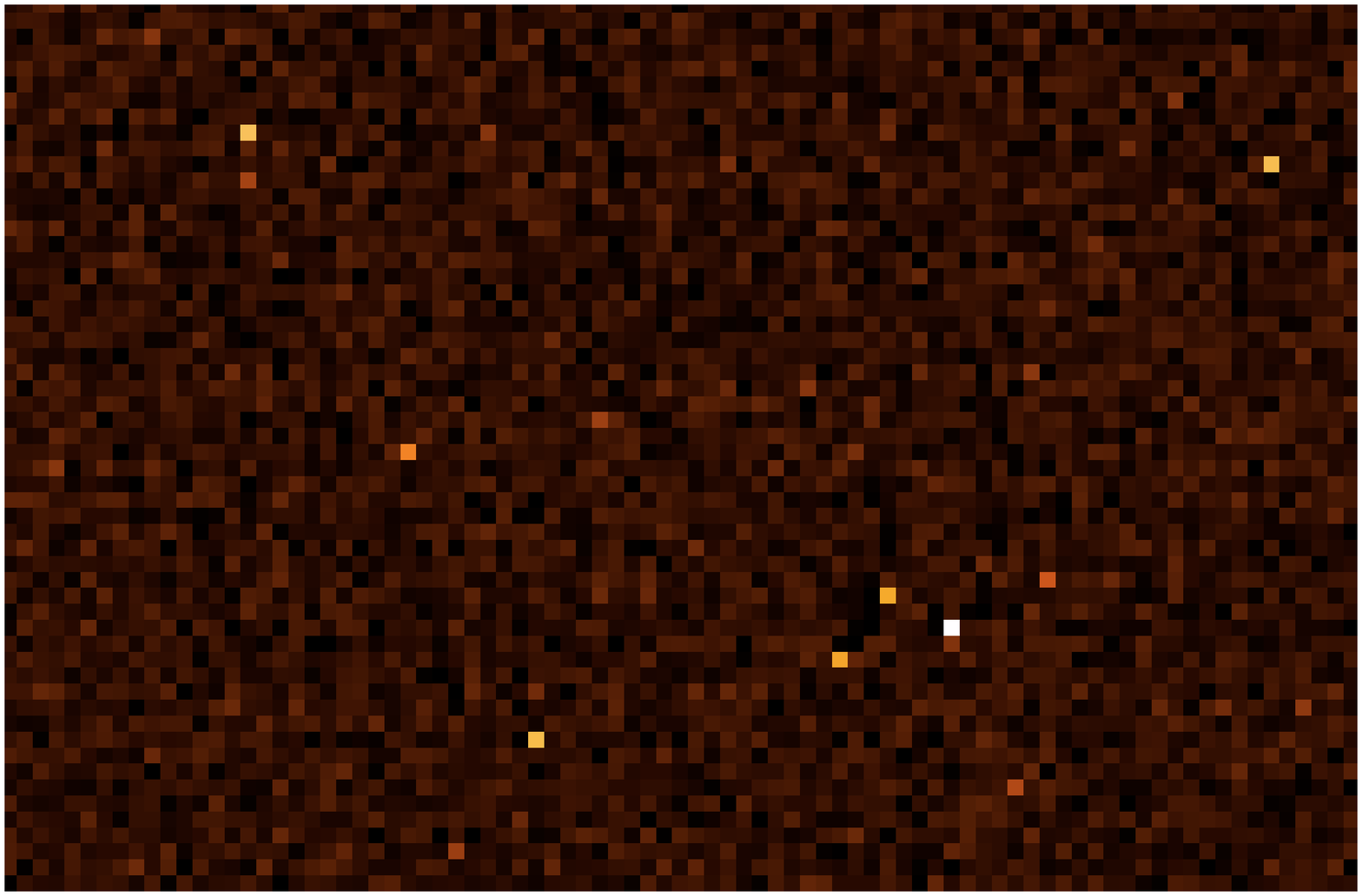}
} {
\includegraphics[width=\figurewidth]{f7a_bw.eps}
\includegraphics[width=\figurewidth]{f7b_bw.eps}
}
\caption{Effect of a bad CTE on EMCCD images taken with a commercial camera and with CCCP/CCD97. Dark images are taken at high gain and were zoomed and false-colored
to enhance details. Both cameras were operated at -85$^\circ$C. \textbf{Left}: Andor camera. We clearly see some pixels that are leaking in the horizontal direction (circled). The event rate (dark+CIC) measured on this camera is 0.0084 electron per pixel per image. The gain over read-out noise ratio is 22. The bad event rate measured is 4.6\% (see text for a definition of bad event rate). \textbf{Right}: CCCP. Pixels leak are far less apparent. One could note that there is also less events since the CIC level is lower in CCCP's images. The event rate (dark + CIC) measured is 0.0018. The gain over readout noise ratio is 22. The bad event rate is 0.3\%.}
\label{fig::realCTE}
\end{center}
\end{figure*}

In order to measure the CIC generated during a read-out of the EMCCD, many ($\sim 1000$) dark frames were acquired. These frames were exposed for the shortest period of time (to minimize the dark noise) and in total darkness\footnote{For CCCP/CCD97, two frame transfers were done before reading-out, yielding a 0-s integration time. However, dark signal is generated during the read-out.}. Then, the histogram of all the frames is fitted with the EM output probability equation (equation \ref{eqn::emOutputProb}), assuming that $n$ is always 1. For very low event rates (in this case, $\simeq$0.003 event pixel$\mathrm{^{-1}}$ per frame), this assumption has little effect on the CIC measurement. This gives at the same time the EM gain, $G$, and the mean quantity of events per pixel. The histogram fitting has also the advantage of seeing all the events that are generated in the image area and storage area and in the conventional horizontal register, even the ones buried in the read-out noise. This is mandatory if the CIC levels are to be compared at different EM gains. Obviously, simply counting the events that are at a 5$\sigma$ threshold would yield lower CIC levels for lower gains as more events would end up in the read-out noise (recall figure \ref{fig::gainEffect}). It is beyond the scope of this paper to describe the algorithm in detail\footnote{Interested readers are invited to go to \url{http://www.astro.umontreal.ca/~odaigle/emccd} to read more about it and see the IDL code.}. This code also calculates the mean bias (CIC+dark free) of all the frames and gives the real read-out noise. The bias calculated by the routine can then be used to remove the bias from light frames.

When a zero integration time is used, or when both the image and storage regions of the CCD are flushed prior to the read-out (dump of the lines through the Dump Gate) the dark signal generated during the readout can further be removed by fitting a slope of the mean signal from the first line read to the last one. This yields the dark count rate per read-out line. This allows the CIC to be completely disentangled from the dark noise.

The Andor camera is advertised as having a CIC+dark level of 0.005 electron pixel$\mathrm{^{-1}}$ per image\footnote{In the datasheet available at \url{https://www.andor.com/download/download_file/?file=L897SS.pdf}} at an EM gain of 1000 for a 30-ms integration time at -85\Deg C. The CIC + dark noise rate measured on such a camera, using the histogram fitting algorithm, is 0.0084. This discrepancy may come in part from the fact that the measurement method used in this article measures all the events instead of counting only the events that are above 5$\sigma$ (which is the way Andor is characterizing its cameras, as stated in their datasheets). This camera is not capable of higher EM gains and it is thus not possible to measure the CIC at higher gains. Obtaining a higher EM gain is just a matter of producing an HV clock that has a higher amplitude. The Andor camera could probably be used at higher EM gain but it is limited by software at 1000. However, as it will be shown in section \ref{sect::gainStability}, the higher the EM gain, the higher the sensitivity of the EM gain to the amplitude of the HV clock. Thus, greater gain must come with greater stability of the HV clock. Moreover, a higher EM gain means a higher CIC level.

Figure \ref{fig::cccpCic} shows the level of CIC that was measured on CCCP/CCD97 at different operating temperatures. At an EM gain of 1000, the amount of events attributable to the CIC is of the order of 0.001 electron pixel$\mathrm{^{-1}}$ per image. At an EM gain of 2500, the CIC level reaches 0.0014 -- 0.0018. This figure also shows that the CIC is not, or is at most weakly, dependent on the operating temperature. Figure \ref{fig::snComparison} compares the SNR of the CCD97 driven by CCCP and the Andor camera, using the EM gain and CIC levels measured.


\subsection{Charge transfer efficiency}
Having a high gain, a low CIC and a low real readout noise are not the only factors that are needed to render the faint flux imaging efficient. The Charge Transfer Efficiency (CTE) of the EM register, that is, the efficiency at which the electrons are moved across the EM register, plays an important role. If this CTE is too low, some of the electrons composing a pixel will not be transferred when they should be and they will lag in the following pixels. In images, these lagging electrons will be seen as a tail lagging behind a bright pixel. Figure \ref{fig::realCTE} shows the effect of an EM CTE that is not optimum (left panel, circled events). These leaking pixels are polluting adjacent pixels by raising their value. They can then be counted as pixels being stroked by a photon when they were not. This creates a source of noise.


For means of comparison, one could define a scenario that would allow these \textit{bad pixels} to be identified and counted. For example, one could use dark frames, expected to count only dark and CIC events. If, say, the rate of events in these images is 0.01 event pixel$\mathrm{^{-1}}$ per image, there is, assuming a Poissonian process for the generation of the dark and CIC, only 1\% chance that an event would immediately follow another. By counting the quantity of events that are adjacent, it is possible to know the quantity of events that are due to the bad CTE.

In the left panel of figure \ref{fig::realCTE}, the mean CIC+dark rate measured is 0.0084 event pixel$\mathrm{^{-1}}$ per frame\footnote{The measurements of the \textit{bad events} involved about a thousand dark frames, in addition to the zoomed region of figure \ref{fig::realCTE}.}. Thus, no more than 0.84\% of the events should be adjacent. However, 5.5\% of the events immediately follow one another. The excess in adjacent events (4.6\%) must come from the bad CTE. In other words, given any event in an image (dark, CIC, or photo-electron), there is a 4.6\% chance that it will be followed by a bad event.

\begin{figure}[tbp]
\begin{center}
\includegraphics[width=\figurewidth]{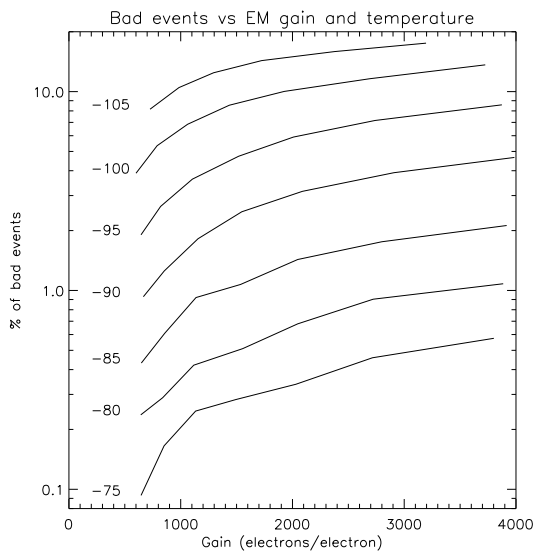}
\caption{Fraction of bad events as a function of the EM gain and temperature (in $^\circ$C), as measured with CCCP/CCD97. The bad event rate shown in this figure is not compensated from the natural rate at which two adjacent events can occur.}
\label{fig::bad}
\end{center}
\end{figure}

In the left panel of figure \ref{fig::realCTE}, the bad event rate is measured to be 0.46\%, where the expected rate is 0.18\%. Thus, one can tell that at -85$^\circ$C and at a gain over read-out noise ratio of 22, there are only $\sim$0.3\% bad events with the CCCP/CCD97 camera. Thus, even if CCCP's EM gain is higher than that of the Andor camera, the first has a much lower bad event rate, which means that the CTE figure of the EMCCD can be better handled with CCCP.

The deterioration of the CTE increases with both the EM gain and the low temperature of the CCD (figure \ref{fig::bad}). Adjustments to the clocking (clock overlap voltages, clock frequency components, clock high and low levels, etc.) were performed after these measurements were made and better CTE figures are now accomplished, as can be seen in figure \ref{fig::realCTE}. However, the behaviour of the CTE as a function of the temperature and gain is maintained: the bad event rate rises with the EM gain and with lower temperatures. 


\subsection{Gain stability}
\label{sect::gainStability}

\begin{figure}[tbp]
\begin{center}
\includegraphics[width=\figurewidth]{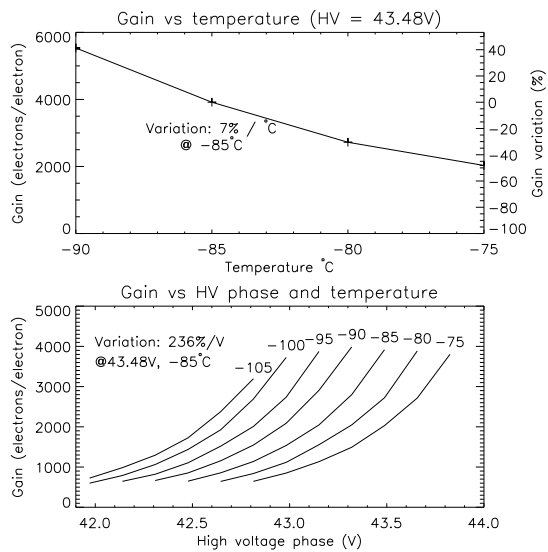}
\caption{\textbf{Top}: Sensitivity of the EM gain to the temperature of the EMCCD. \textbf{Bottom}: Sensitivity of the EM gain to the maximum level of the HV clock.}
\label{fig::gainSensitivity}
\end{center}
\end{figure}

The stability of the gain of an EMCCD is important as it will affect the photometric measurements. The gain of the EM register is very sensitive to the variations of the HV clock. The higher the gain, the higher the sensitivity to the HV clock amplitude (figure \ref{fig::gainSensitivity}, bottom panel). At a high gain, a variation of 1 volt in the amplitude of the HV clock represents more than a two-fold variation of the EM gain. This means that, in order to achieve a gain stability of the order of a percent, the HV clock must be stable within $\sim$5 mV. The EM register is also sensitive to the temperature (figure \ref{fig::gainSensitivity}, top panel). In the EM gain regime shown, the temperature must be controlled within $\pm$0.1\Deg C in order to achieve a gain stability of $\pm$1\%. 

\subsubsection{Gain stability over time}
\label{sect::emGainStabilityOverTime}

The stability of the EM gain of CCCP/CCD97 has been measured by taking dark images continuously at high gain ($G/\sigma \simeq 30$), moderate frame rate (10 fps), and at a temperature of -85\Deg C. In total, 30000 images were acquired. Then, the EM gain of these images was determined using the usual algorithm. The algorithm used a window of 400 frames out of the 30000 and slid 400 frames every iteration. Thus, 75 data points were produced, each representing about 40 s. The results are shown in figure \ref{fig::gainStability}, left panel. The error bars in this figure are calculated using
\begin{equation}
\label{eqn::errorGain}
e = G\sqrt{\dfrac{2}{n}},
\end{equation}
where $G$ is the EM gain measured and $n$ is the quantity of events that were used to obtain $G$. The constant $2$ is used to account for the ENF of the EMCCD.

The gain variation over time is thought to be due to the variation of the temperature, rather than the variation of the HV clock amplitude. The temperature controller used to gather these data has an accuracy slightly worse than $\pm$0.1\Deg C.  


\subsubsection{Gain stability over frame}
The EM gain must also be stable through an image or, at least, its variation must be well characterized. EM gain variation may come from the HV clock amplitude which can take some time to stabilize at the beginning of the frame, given that the HV clock is turned off during the integration time. Even if the HV clock is kept running during the integration time, the capacitive coupling between it and the conventional horizontal clocks can induce a variation of the HV amplitude when they are activated and, consequently, a variation in the gain.

\begin{figure*}[tbp]
\begin{center}
\includegraphics[width=\figurewidth]{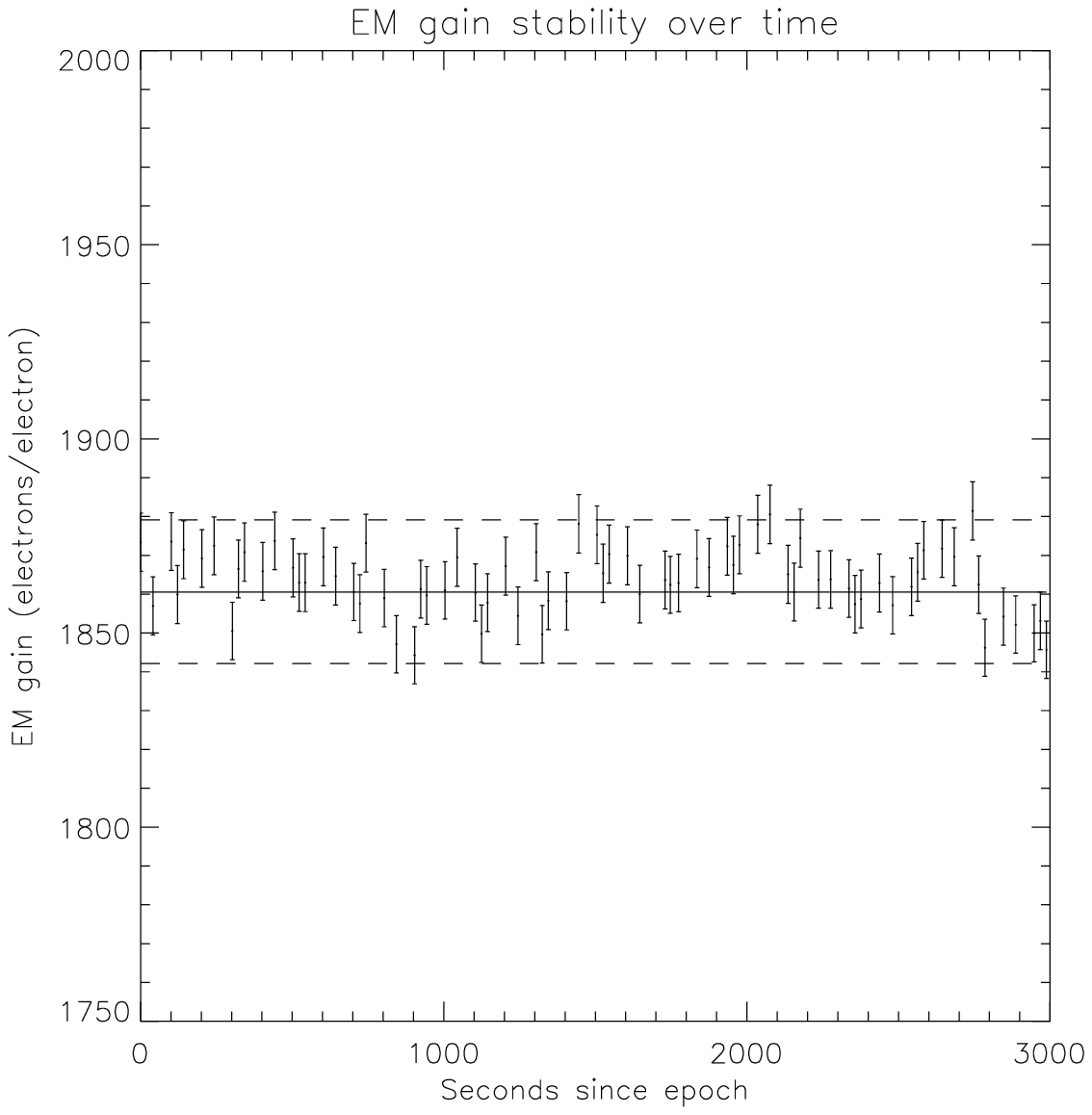}
\includegraphics[width=\figurewidth]{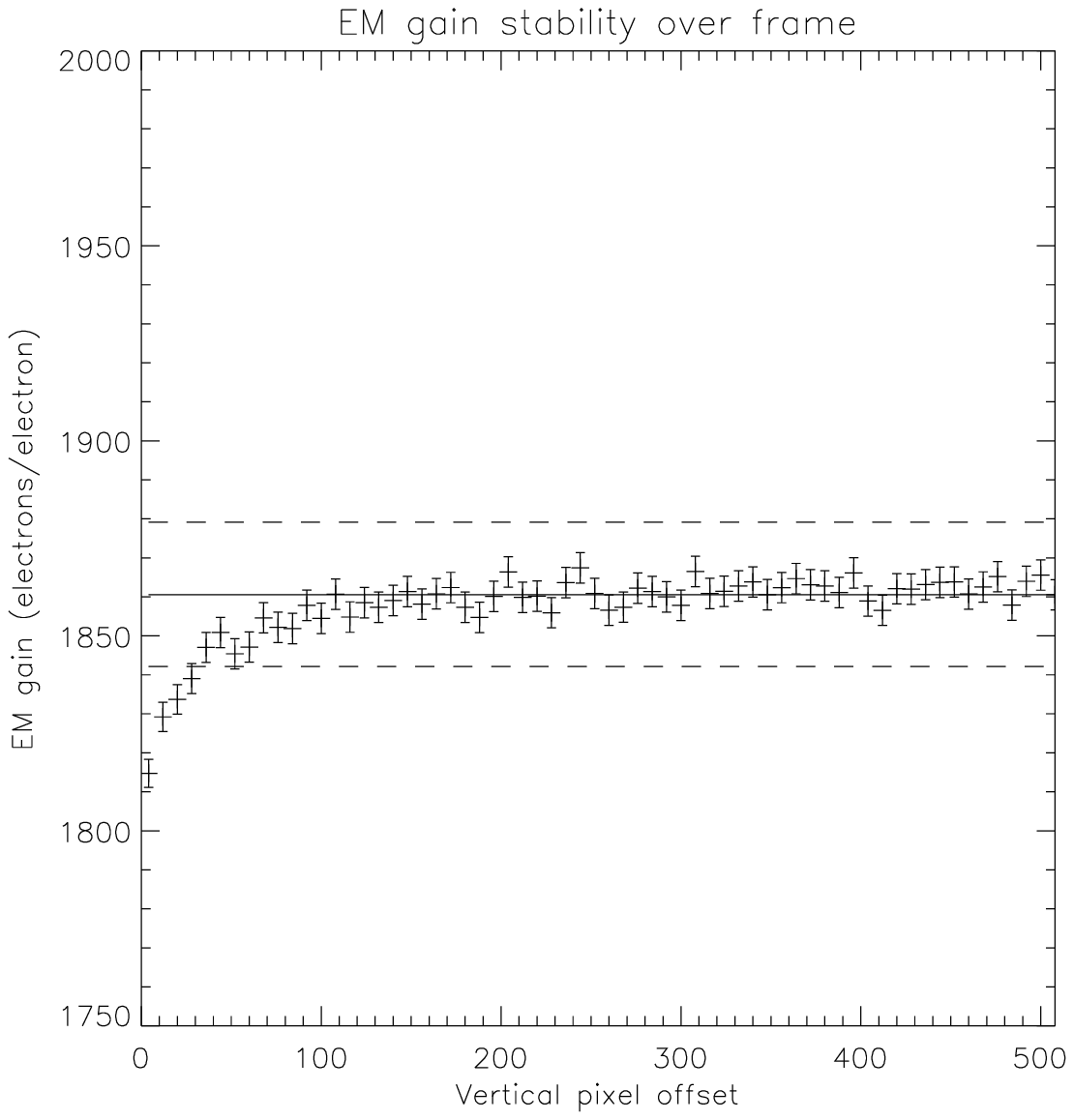}
\caption{Stability of the EM gain. The plain line shows the median of the data set while the dashed lines shows a +/- 1\% variation. \textbf{Left}: Stability over time. \textbf{Right}: Stability over frame.}
\label{fig::gainStability}
\end{center}
\end{figure*}

The 30000 frames used in section \ref{sect::emGainStabilityOverTime} were used once again to measure the gain over the image. The EM gain was determined on a per-line basis rather than on an image basis. Then, lines were binned by 8 to increase the SNR (given equation \ref{eqn::errorGain}). This yields the right panel of figure \ref{fig::gainStability}.

One can see that the HV clock takes about 32 lines to stabilize within the $\pm$1\% gain variation. The HV clock was stopped during the exposure time of these frames. This gain variation is stable, which can be taken into account while processing the raw data to yield accurate photometric measurements.

\subsection{Experimental SNR}
\label{sect::experimentalSNR}
Experimental SNR curves were obtained in the lab. In order to do so, a scene spanning about 3 orders of magnitude in contrast has been observed (figure \ref{fig::experimentalSn}, bottom panels). Two acquisitions were performed: PC and AM, where the exposure time was set to 0.05s and 0.5s, respectively.

Even though care has been taken to put the lab in total darkness, the background of the scene was still very faintly lit (about 0.02 photon pixel$\mathrm{^{-1}}$ s$\mathrm{^{-1}}$), which explains the background signal. In total, 86000 images with an exposure time of 0.05s and 8600 images with an exposure time of 0.5s were acquired. The same amount of dark frames having the same exposure time were also acquired to remove the CIC+dark component. The dark frames were used to calculate the effective EM gain, by histogram fitting (see section \ref{sect::cccpCic}). The mean bias has been extracted from the dark frames and has been used to subtract the bias of all the light frames.

As for any CCCP data acquisition, the raw data coming out of CCCP during this experiment were stored in fits files during the acquisition and these files are then processed off-line. The resulting SNR curves are presented in figure \ref{fig::experimentalSn}. The frames where a cosmic ray was detected were simply removed. This accounted for about 0.15\% of the frames in PC and for about 1.5\% in AM. 

\subsubsection{PC processing}
The PC frames were processed with a threshold of $5\sigma$. However, the sum of the PC frames, minus the dark frames, gives only the amount of counted photons, $c_p$, whereas one wants to know the sum of incident photons, $f$. There are two possible sources of losses:
\begin{itemize}
\item The threshold, which is responsible for the loss of the events not amplified enough and ending up in the read-out noise;
\item The events that are lost due to coincidence.
\end{itemize}

The proportion of the events lost due to the threshold, $e_l$, is given by equation \ref{eqn::gainEffect} if one uses $c_p$ as $\lambda$. The values of $G$ and $cut$ were, in this case, $30\sigma$ and $5\sigma$, respectively. Thus, the correction can be made simply by dividing the $c_p$ by $1-e_l$. This gives the amount of detected photons, $d_p$.

The coincidence losses can be modelled by simple Poissonian statistics. If the amount of incident photons, $f$, is unknown, the probability of having zero photon is known. This is simply $1-d_p$. Thus, $1-d_p$ represents the probability of having zero photon under a flux of $f$, and $f$ can be recovered by means of
\begin{equation}
\label{eqn::coincidenceLossesCorrection}
f = -\ln{(1-d_p)}.
\end{equation}
This correction, however, induces an excess noise that scales as
\begin{equation}
F_c = \dfrac{1}{e^{-f}}.
\end{equation}
At low flux, $F_c$ approaches the value of 1. It is only under moderate fluxes, where coincidence losses become important, that $F_c$ prevails.

Now, one can calculate the effective SNR of the image for every pixel. The SNR is given by
\begin{equation}
SNR_{PC} = \dfrac{f}{\sqrt{F_c^2 N}},
\end{equation}
where $N$ is the variance of the pixel over all the 86000 frames. This SNR curve can then be compared to the curve of a perfect PC system whose noise would be only the shot noise, which is $SNR = \sqrt{f}$. This gives the experimental points shown in figure \ref{fig::experimentalSn}, left panel. The theoretical curve is given by
\begin{equation}
SNR_{PC_t} = \dfrac{S}{\sqrt{(S+D)g}},
\end{equation}
where $S$ is the incident flux, and $D$ is the mean CIC+dark rate (the mean of all the dark frames). The term $g$ is the excess noise induced by the coincidence loss and its correction that are implied by the PC processing:
\begin{equation}
g = \dfrac{S+D}{e^{-(S+D)}}.
\end{equation}
In this case, $D$ is \cic\, event pixel$\mathrm{^{-1}}$ per image. Of course, all of these corrections do not take into account the photons lost due to the QE. However, there is nothing CCCP can do about it and the plots are made according to a perfect photon counting system that would have the same QE as the CCD97.

The agreement between the theoretical and the experimental curves for the PC processing is nearly perfect (figure \ref{fig::experimentalSn}, top left panel). The image resulting from the averaging of all 86000 frames is shown in the bottom left panel of the figure.

\subsubsection{AM processing}
\begin{figure*}[tbp]
\begin{center}
\includegraphics[width=\figurewidth]{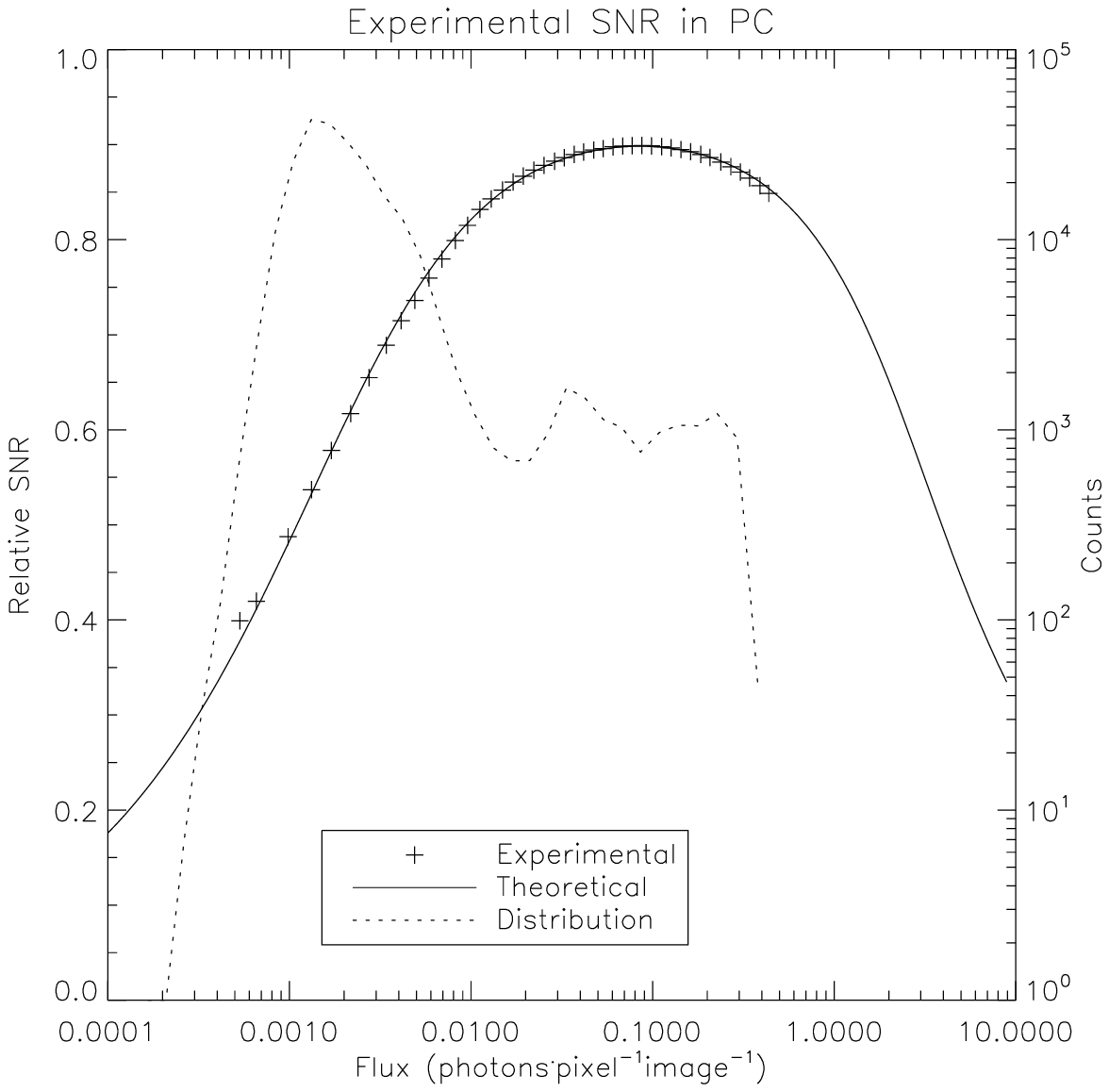}
\includegraphics[width=\figurewidth]{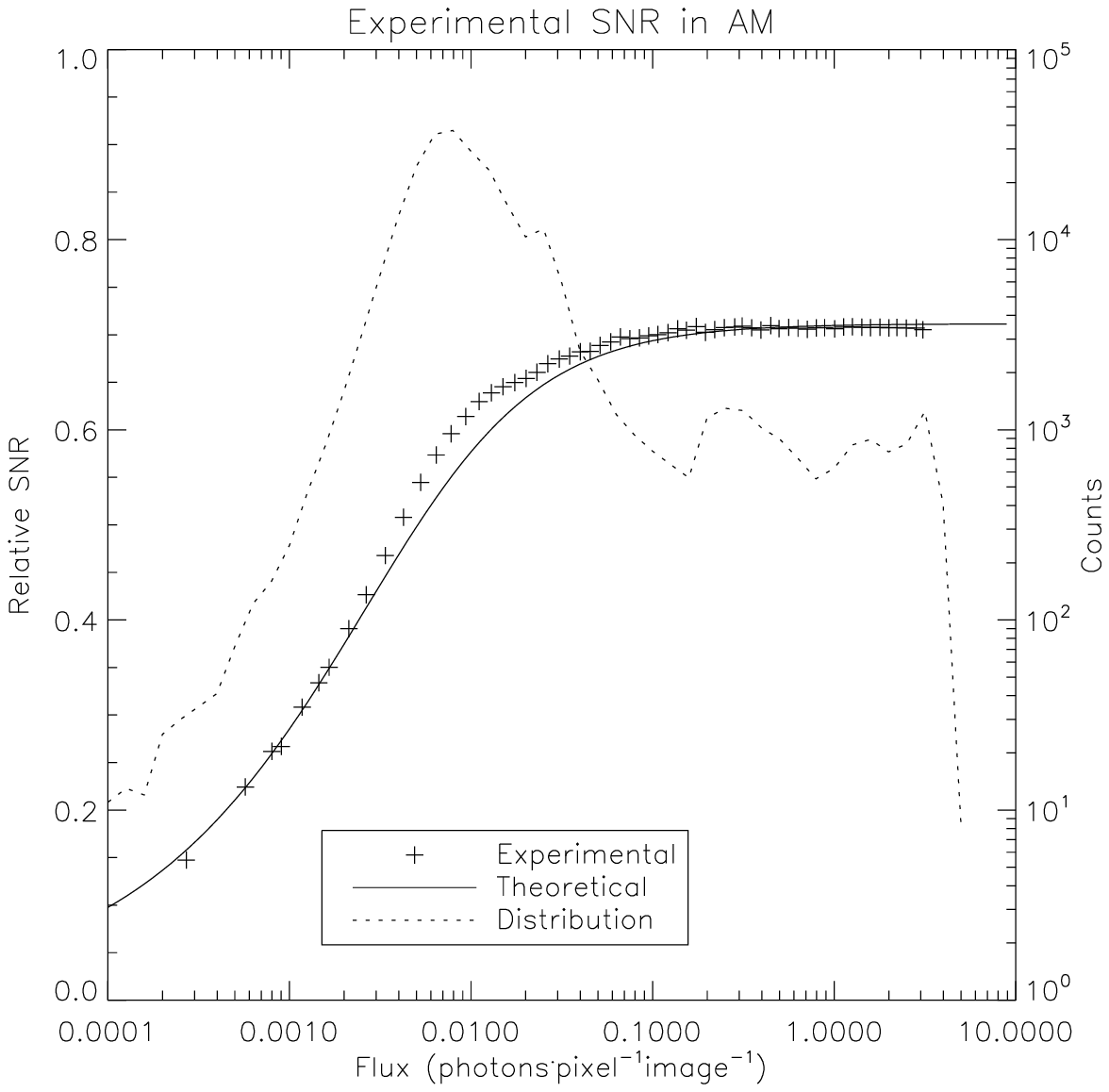}
\ifthenelse{\colorFigures=1} {
\includegraphics[angle=270, width=\figurewidth]{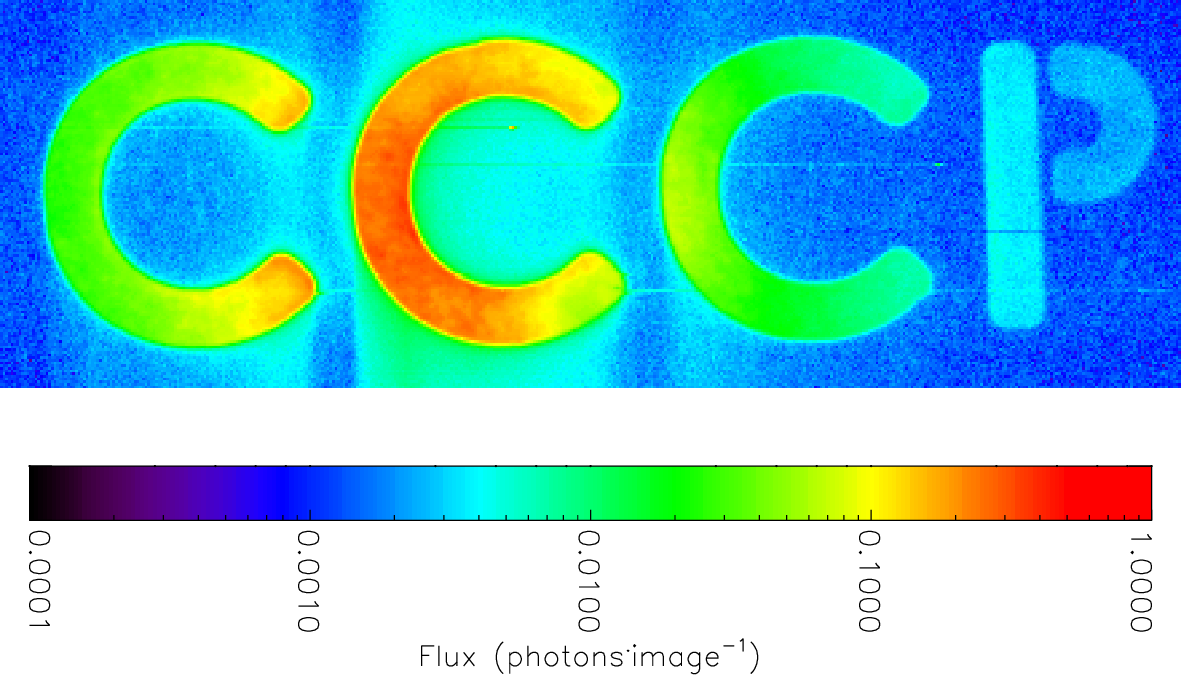}
\includegraphics[angle=270, width=\figurewidth]{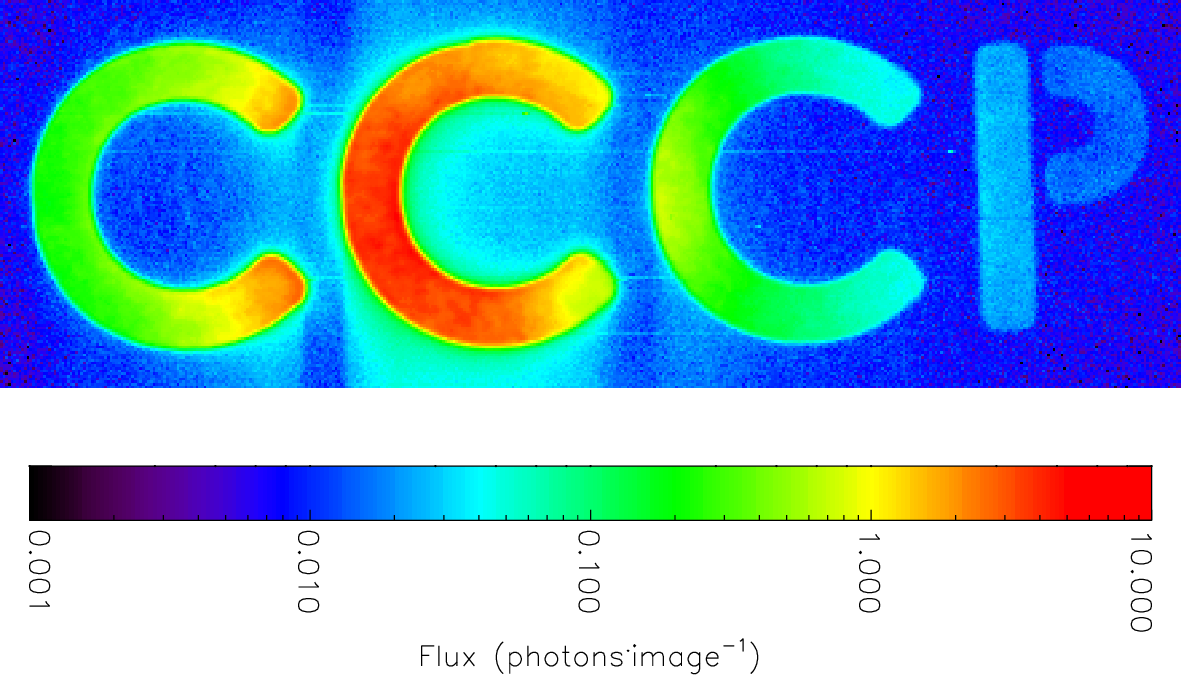}
}{
\includegraphics[angle=270, width=\figurewidth]{f12c_bw.eps}
\includegraphics[angle=270, width=\figurewidth]{f12d_bw.eps}
}
\caption{\textbf{Top}: Experimental SNR curves. Curves were obtained by acquiring 80000 images with CCCP/CCD97 of a very low lit scene. The exposure time was 0.05s and 0.5s for the PC and AM acquisitons, respectively. Both acquisition were made at a $G/\sigma$ ratio of 30. The theoretical plots are those of an EMCCD having the same $G/\sigma$ ratio and a CIC+dark rate of \cic\, and \cicAm\, for PC and AM, respectively. A pixel value distribution is superimposed on the plots. \textbf{Left}: SNR after PC processing. \textbf{Right}: SNR after AM processing. \textbf{Bottom}: Mean of all the images acquired to produce the plots of the top pannels, dark+CIC subtracted. The images are rotated by 90\Deg. The horizontal shifting direction is from top to bottom and the vertical shifting direction is from right to left. \textbf{Left}: Image of the PC acquisition, after PC processing. \textbf{Right}: Image of the AM acquisition, after AM processing.}
\label{fig::experimentalSn}
\end{center}
\end{figure*}

The case of the AM processing is simpler. The flux in a pixel is given by dividing its value by the EM gain at which the image was acquired. The mean flux of a pixel, $f$, is simply given by the mean value of the same pixel across all the images minus the dark signal measured for that pixel. The SNR of a pixel is simply given by
\begin{equation}
SNR_{AM} = \dfrac{f}{\sqrt{N}}, 
\end{equation}
where $N$ is the variance of that pixel through all of the AM frames. When this SNR curve is normalized by that of a perfect photon counting system, this gives the experimental curve shown in figure \ref{fig::experimentalSn}, right panel.

The theoretical curve of that figure is obtained by taking into account the ENF, $F$, that plagues the EMCCD. Thus, the SNR equation is given by
\begin{equation}
SNR_{AM_t} = \dfrac{f}{\sqrt{(f^2+d^2)F^2 + \sigma_{eff}^2}},
\end{equation}
where $\sigma_{eff}$ is the effective read-out noise. Since the EM gain is high (30$\sigma$), $F^2$ takes the value of 2. The CIC+dark, $d$, is \cicAm\, in that plot.

In order to reach very faint fluxes with the AM processing, the bias needs to have an extremely high SNR. There must be as many frames used to build the bias as there are light frames to process. If not, since the same bias is used for all the light frames, the noise of the bias will become the dominant source of noise in the final images. The PC processing is not as sensitive as the AM processing to the SNR of the bias. In PC, the bias's noise will be removed by the threshold applied to the data.

The plots also shows that the EM gain is accurately determined, which is critical for AM processing. The experimental AM plot would shift up or down if the EM gain were underestimated or overestimated, respectively.


\subsubsection{Comments}
A result of this experiment that might have struck an attentive reader is the higher than expected CIC+dark rate in PC. On measurements made with histogram fittings (as in figure \ref{fig::cccpCic}), the CIC+dark rate was about 0.0018 for a $G/\sigma$ ratio of $\sim$30. The fitting of the SNR curves needed a CIC+dark level of 0.0023 in order to agree with the experimental data. This is also the mean event rate measured in the dark frames of the PC acquisition. This behaviour is explained through CIC and dark electrons generated into the EM register. This is in agreement with the fact that the mean source of CIC in CCCP/CCD97 comes from the CIC that is generated into the horizontal register \citep{2008SPIE.7014E.219D}. Given that the EM register sole existence is driven by the fact that it generates electrons, it is understandable that it generates CIC too. Dark electrons can also be generated in the EM register. The mean amplification of these CIC and dark electrons varies according to the elements into which the charge is created. The exact amount of CIC that is generated into the EM register is hard to compare at different $G/\sigma$ ratios. At low $G/\sigma$, mostly all of these CIC events will end up in the read-out noise, making it impossible to count them. So, it has been decided not to count them when comparing CIC levels. However, they must be accounted for in the SNR plots. 

The effect of these lightly multiplied events will be stronger in the PC data than in the AM data. The PC-processed data see an event as an event, regardless of its multiplication factor. The AM-processed data, however, will see, on average, these events as having less than one input electron. Thus, the CIC+dark level computed in PC will differ from the one computed in AM. As an example, the CIC+dark level of the PC dark frames (0.05s exposure time) of figure \ref{fig::experimentalSn} is \cic. If these dark frames are processed in AM, the mean signal level measured is 0.0015 event/pixel/image, which is closer to the values presented in figure \ref{fig::cccpCic}. Given these facts, the threshold used for the PC processing could be raised (6$\sigma$, 7$\sigma$) to avoid counting some of theses events. Of course, the proportion of counted photons would diminish. An optimum threshold, yielding the best SNR, could be calculated.

\begin{figure*}[tbp]
\begin{center}
\ifthenelse{\colorFigures=1} {
\includegraphics[angle=270, width=\figurewidth]{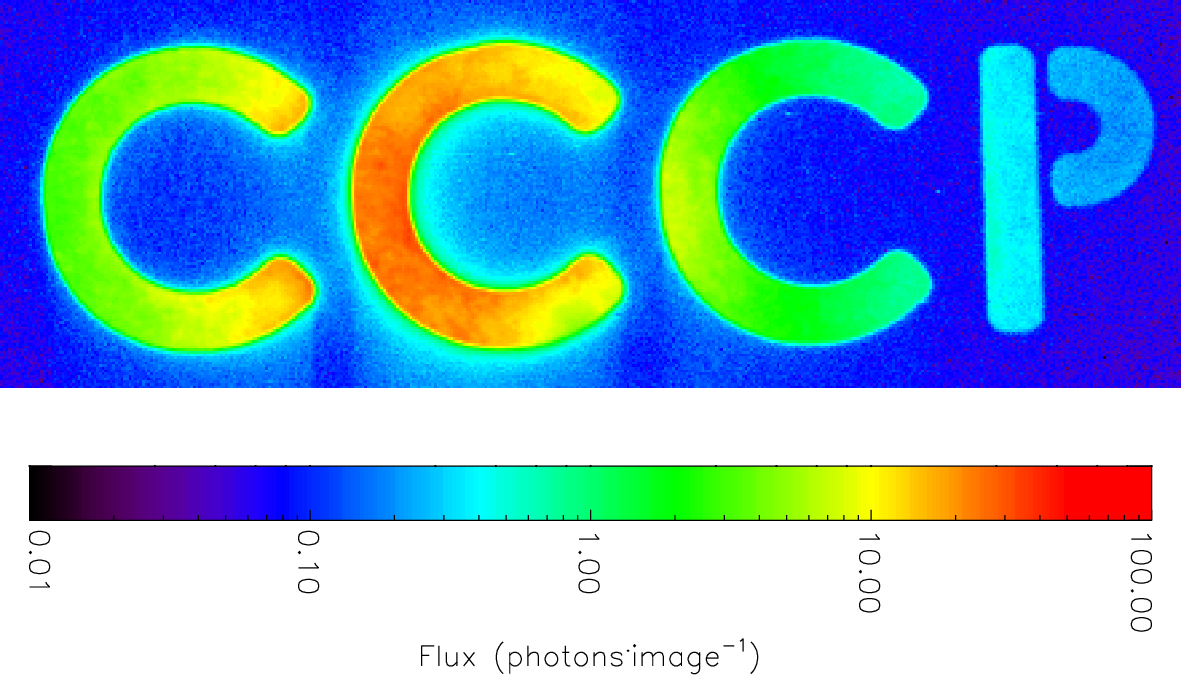}
\includegraphics[angle=270, width=\figurewidth]{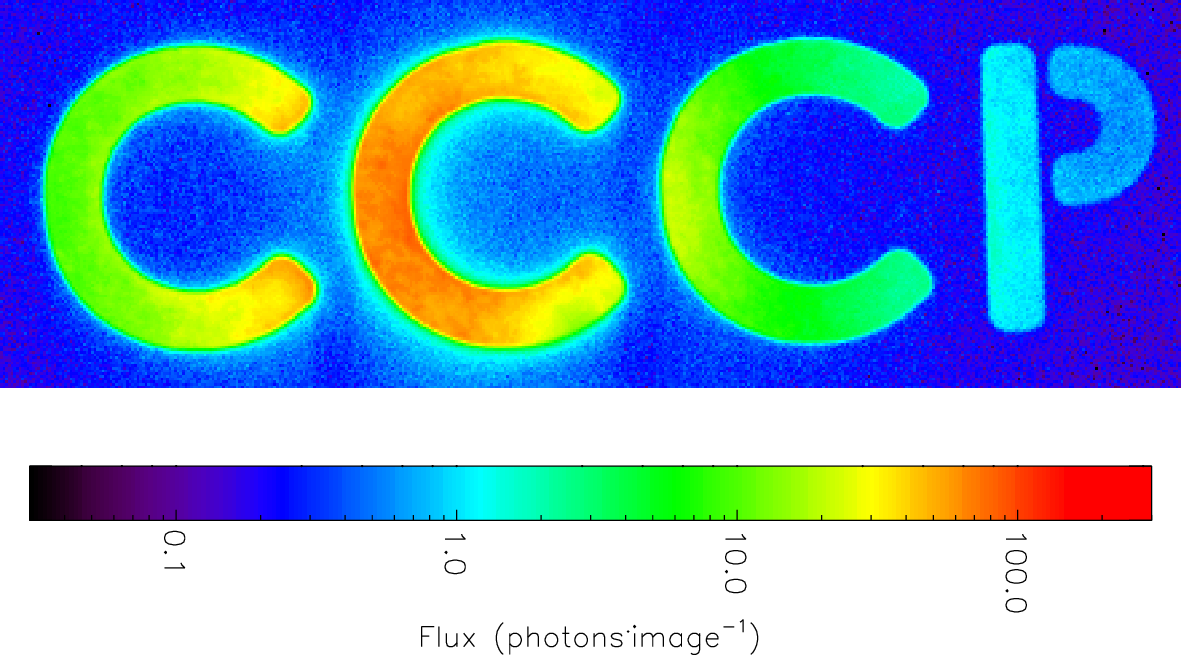}
} {
\includegraphics[angle=270, width=\figurewidth]{f13a_bw.eps}
\includegraphics[angle=270, width=\figurewidth]{f13b_bw.eps}
}
\caption{\textbf{Left}: Same as the AM image of figure \ref{fig::experimentalSn}, but with an exposure time of 5s per frame. \textbf{Right}: Same as the left panel, but with an exposure time of 15s per frame.}
\label{fig::highFlux}
\end{center}
\end{figure*}

The image of the low light scene (figure \ref{fig::experimentalSn}, bottom panel) shows trailing electrons behind the most luminous pixels. The trail is about two orders of magnitude fainter than the mostly lit pixel on its line. This could be caused either by the low horizontal CTE in the EM register or the heating of the multiplication register when many electrons are generated into it. It is unlikely that this trailing happens in the conventional horizontal register as there is very rarely more than one electron per element in the PC image (the highest mean flux being about 0.5 photon pixel$\mathrm{^{-1}}$ per image). Surprisingly, this effect is less visible in the AM image, where more electrons/pixel are gathered, given the 10-fold increase in exposure time. This relation hold for even longer exposure times, as shown in figure \ref{fig::highFlux}, where exposure times of 5 and 15 s frame$\mathrm{^{-1}}$ were used, at the same EM gain (AM processing). These frames also show that these trailing electrons are not the result of an overflow of the horizontal register since it would worsen as the exposure time increases and more electrons are gathered. This phenomenon, however, did not compromise the quality of the scientific data, as discussed in section \ref{sect::scientificResults}.

\section{Scientific results}
\label{sect::scientificResults}

CCCP/CCD97 was tested at the focal plane of the FaNTOmM integral field spectrograph \citep{2002PASP..114.1043G, 2003SPIE.4841.1472H}. FaNTOmM is basically a focal reducer and a narrow band interference filter ($\sim10$\AA) coupled to a high resolution ($ R > 10000$) Fabry-Perot (FP) interferometer. FaNTOmM falls into the same instrument category as the newer G\ha FAS \citep{2008ASPC..390..168C, 2008PASP..120..665H}. The FaNTOmM instrument is mostly used to map the kinematics of galaxies using the doppler shift of the \ha\, line (such as in \cite{2005MNRAS.360.1201H, 2006MNRAS.366..812C, 2006MNRAS.367..469D, 2008MNRAS.385..553D, 2008MNRAS.388..500E}). The \ha\, emission comes from both luminous \hII\, regions and faint, diffuse \ha\, regions. There are a few strong sky emission lines caused by the OH radicals around \ha\, but there are many dark regions in the spectrum as well. Given the high spectral resolution of these observations, they are mostly read noise limited instead of sky background limited. Moreover, since a FP interferometer is a scanning instrument, it is of great interest to scan the interferometer many times throughout an observation to average the photometric variations of the sky. This requires many short exposures (typically 5-15 seconds between the moves of the interferometer). These kind of observations are perfectly suited for a photon counting camera and this is the reason why FaNTOmM is usually fitted with an IPCS as the imaging device.

The mean drawback of the IPCS is its low QE, which is $\sim20$\% for the case of the IPCS of FaNTOmM. However, the IPCS have a very low dark noise, typically of the order of $10^{-5}$ electron pixel$\mathrm{^{-1}}$ s$\mathrm{^{-1}}$. The advantage of the EMCCD is obvious in terms of gain in QE if the CIC of the EMCCD is low enough \citep{2004SPIE.5499..219D}. The results obtained in lab with CCCP/CCD97 were promising. Engineering telescope time was obtained in September 2008 on the 1.6-m telescope of the Observatoire du mont M\'egantic to test CCCP on real-world objects. During this run, the CCCP/CCD97 camera was used at the focal plane of the FaNTOmM instrument, in place of the IPCS.

\subsection{Observations}

The galaxy \mbox{NGC 7731} was observed during this run. This galaxy has already been observed with FaNTOmM/IPCS through the SINGS \ha\, survey \citep{2006MNRAS.367..469D}. This observation will allow the comparison of the sensitivity of CCCP/CCD97 with FaNTOmM/IPCS.

The parameters of the observations are presented in table \ref{table::FPobservations}. All the data cubes were processed with the data reduction techniques presented in \cite{2006MNRAS.368.1016D}. The sky emission was subtracted by a sky cube fitting. No spatial smoothing was applied. The Radial Velocity (RV) maps were extracted and then cleaned by an automatic routine that correlates a pixel to the continuum and monochromatic flux-weighted value of the neighbouring ones to determine its validity, based on a maximum deviation. Then, a manual clean-up removes the pixels whose RV has been incorrectly determined. This is usually due to some sky residuals that are polluting the pixel.

\begin{table*}
\caption{Parameters of the \ha\, Fabry-Perot observations of \mbox{NGC 7331}.} 
\begin{center}
\begin{tabular}{ l  m{0.18\textwidth}  m{0.18\textwidth}  m{0.18\textwidth}}
\hline \hline
Observation name & IPCS & AM & PC \\ \hline
Detector & GaAs IPCS & CCCP/CCD97 & CCCP/CCD97 \\ \hline
Geometry & 512 x 512 & 512 x 512 & 512 x 512 \\ \hline
Pixel size (\Sec) & 1.6 & 1.07 & 1.07 \\ \hline
QE at \ha\, (\%)& $\sim$ 20 & $\sim$ 90 & $\sim$ 90 \\ \hline
$\sigma/G$ ratio & N/A & 30 & 30 \\ \hline
CIC ($\mathrm{\bar{e}}\cdot \mathrm{frame}^{-1}$) & N/A & 0.0015 & 0.0022 \\ \hline
Dark current ($\mathrm{\bar{e}}\cdot \mathrm{s}^{-1}$) & 0.00001 & 0.0012 & 0.0012 \\ \hline
Processing & PC & AM & PC \\ \hline
Date (UT) & 2002-11-03 & 2008-09-11 & 2008-09-11 \\ \hline
Time (UT) & 23:15:00 & 00:30:00 & 04:15:00 \\ \hline
Telescope & OMM 1.6-m & OMM 1.6-m & OMM 1.6-m\\ \hline
Exterior T\Deg\, (\Deg C)& -12 & 3.5 & 3.5 \\ \hline
Moon age (days)& 27 & 11 & 11 \\ \hline
Moon illumination (\%)& 3 & 90 & 90 \\ \hline
Sky conditions & Mostly clear & Cirrus towards the end of the integration & Cirrus at the beginning of the integration \\ \hline
Mean flux / cycle &
\includegraphics[width=0.15\textwidth]{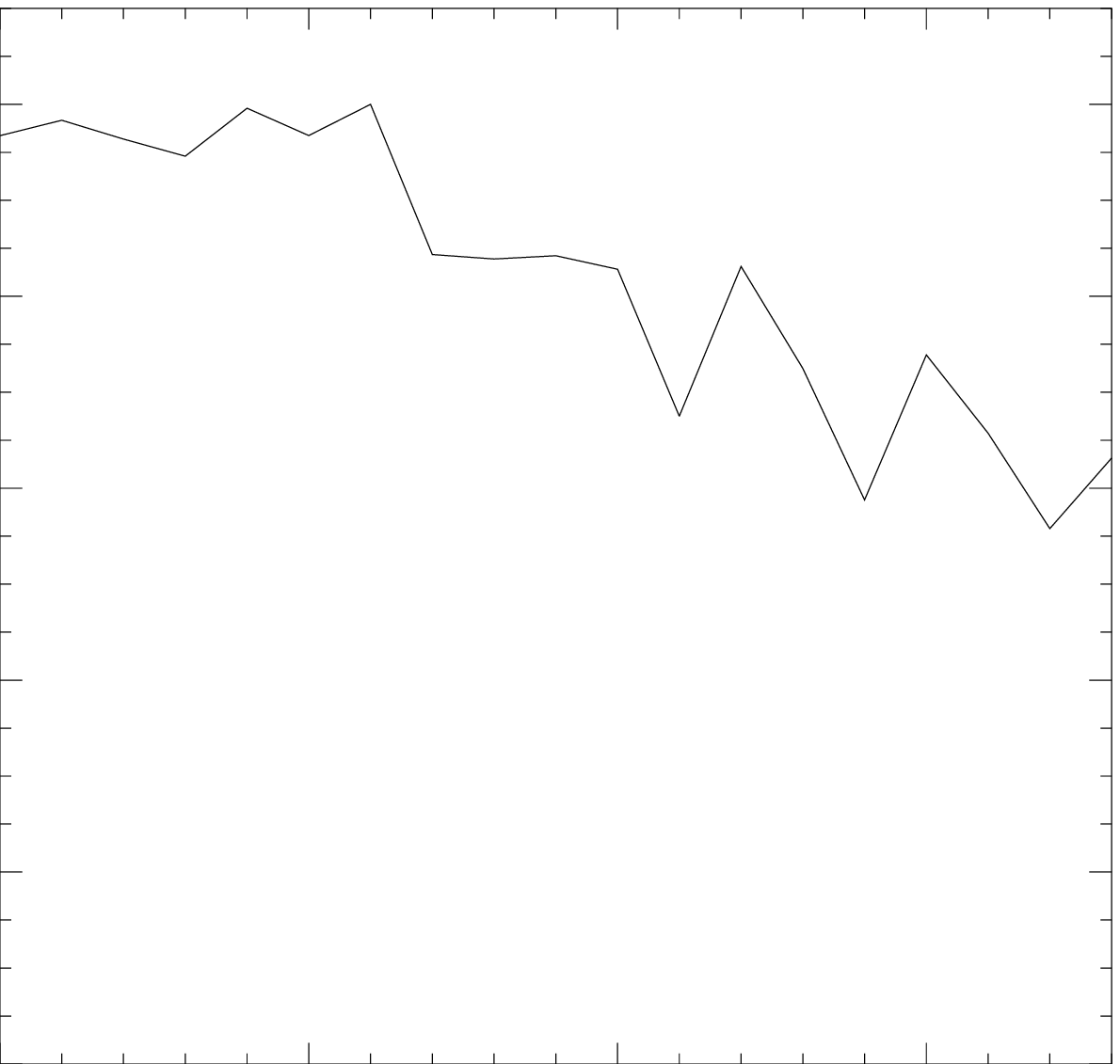} &
\includegraphics[width=0.15\textwidth]{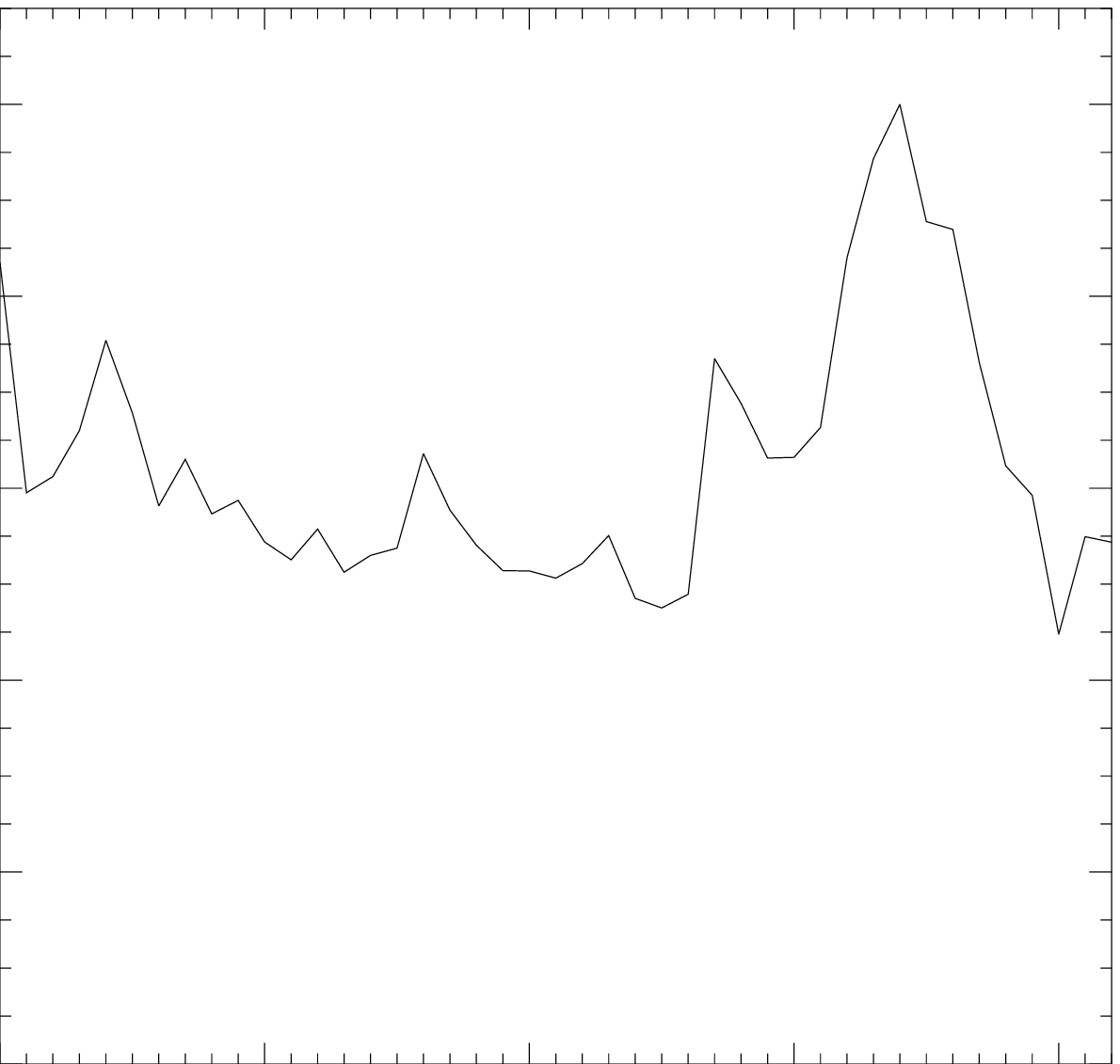} &
\includegraphics[width=0.15\textwidth]{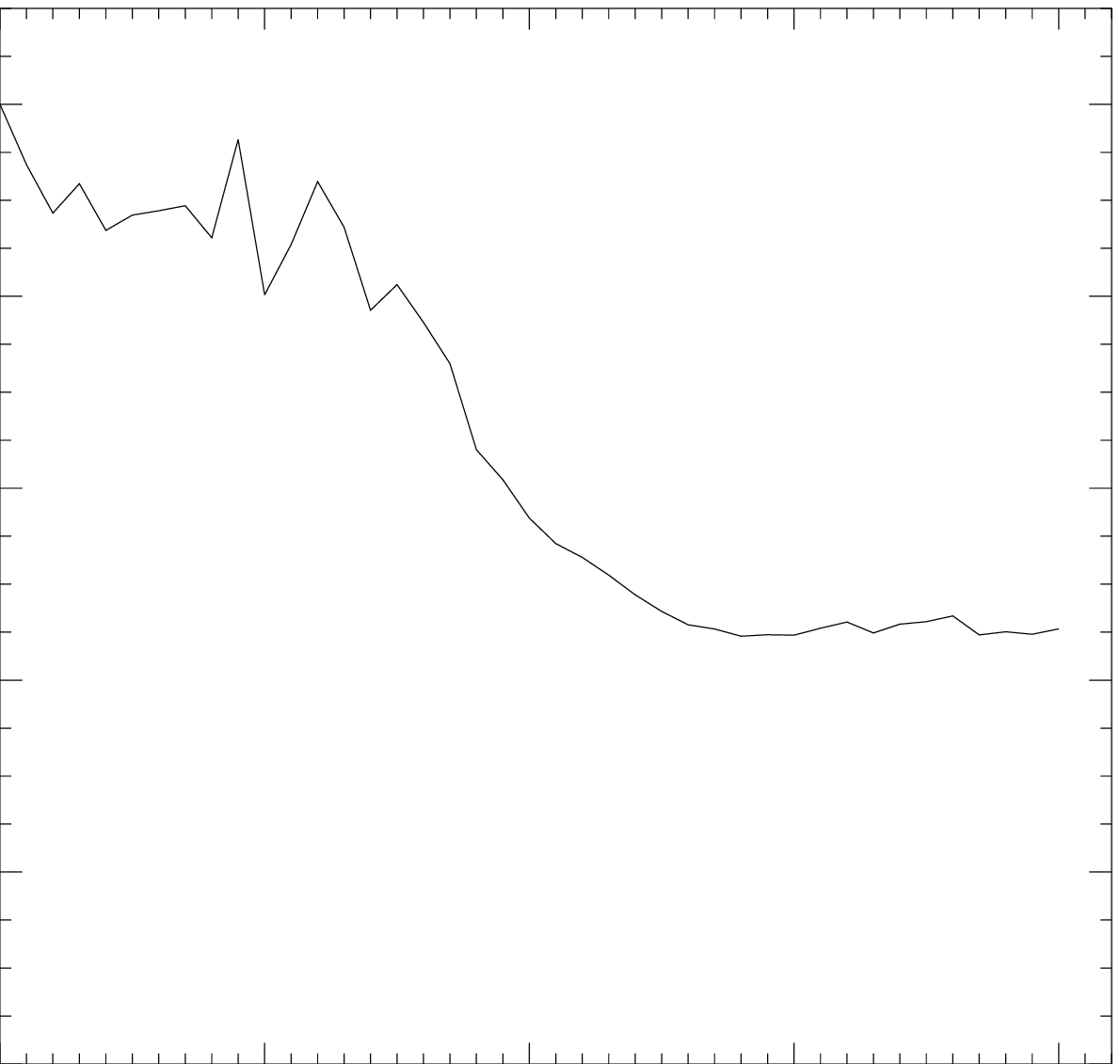}
\\ \hline
Integration time / image (s) & 0.025 & 5 & 0.5\\ \hline
Images / channel & 400 & 1 & 10 \\ \hline
Integration time / channel (s) & 10 & 5 & 5\\ \hline
Number of channels & 48 & 48 & 48 \\ \hline
Number of cycles & 19 & 43 & 41\\ \hline
Total integration time (min)& 152 & 172 & 164 \\ \hline
Filter name & M3 & M3 & M3\\ \hline
Filter $\lambda_{center}$ at outdoor T\Deg\, (\AA)& 6578.4 & 6580.9 & 6580.9 \\ \hline
Filter FWHM (\AA)& 15.5 & 15.5 & 15.5 \\ \hline
FP order at \ha& 765 & 765 & 765\\ \hline
\hline
\end{tabular}
\end{center}
\begin{flushleft}
\end{flushleft}
\label{table::FPobservations}
\end{table*}

In table \ref{table::FPobservations}, the plot of the mean flux per cycle is shown. This, coupled to the moon and sky conditions, gives a rough idea of the quality of the data. The moon was absent when the observation with the IPCS was made and the sky was mostly clear. Thus, the slight drop in the flux towards the end of the observation can be explained by absorption. On the other hand, the moon was mostly full (hence the engineering time) during the CCCP/CCD97 observations and there were a few cirrus. The cirrus were lit by the moon and this caused the flux to rise towards the end of the AM and at the beginning of the PC observation. Moreover, the moon's light excites the OH radicals in the upper atmosphere and makes them glow brighter. This causes the sky emission lines around \ha\, to be stronger, which, once subtracted, leaves a noisier background due to shot noise. The same thing happens once the brighter foreground caused by the clouds lit by the moon is subtracted. It thus takes more photons from the galaxy to reach a given SNR in the CCCP/CCD97 observations.

On the other hand the interference filter's transmission also changes from the IPCS to the CCCP/CCD97 observation because of the exterior temperature. Given its systemic velocity of 816 km s$\mathrm{^{-1}}$, \mbox{NGC 7331}'s rest \ha\, wavelength is at 6580\AA. Its \ha\, emission was better centred in the CCCP/CCD97 observation. On the IPCS observation, this should benifit the approaching side of the galaxy while weakening the receding side.

\subsection{IPCS and CCCP/CCD97 comparisons}

\begin{figure*}[tbp]
\begin{center}
\includegraphics[width=\textwidth]{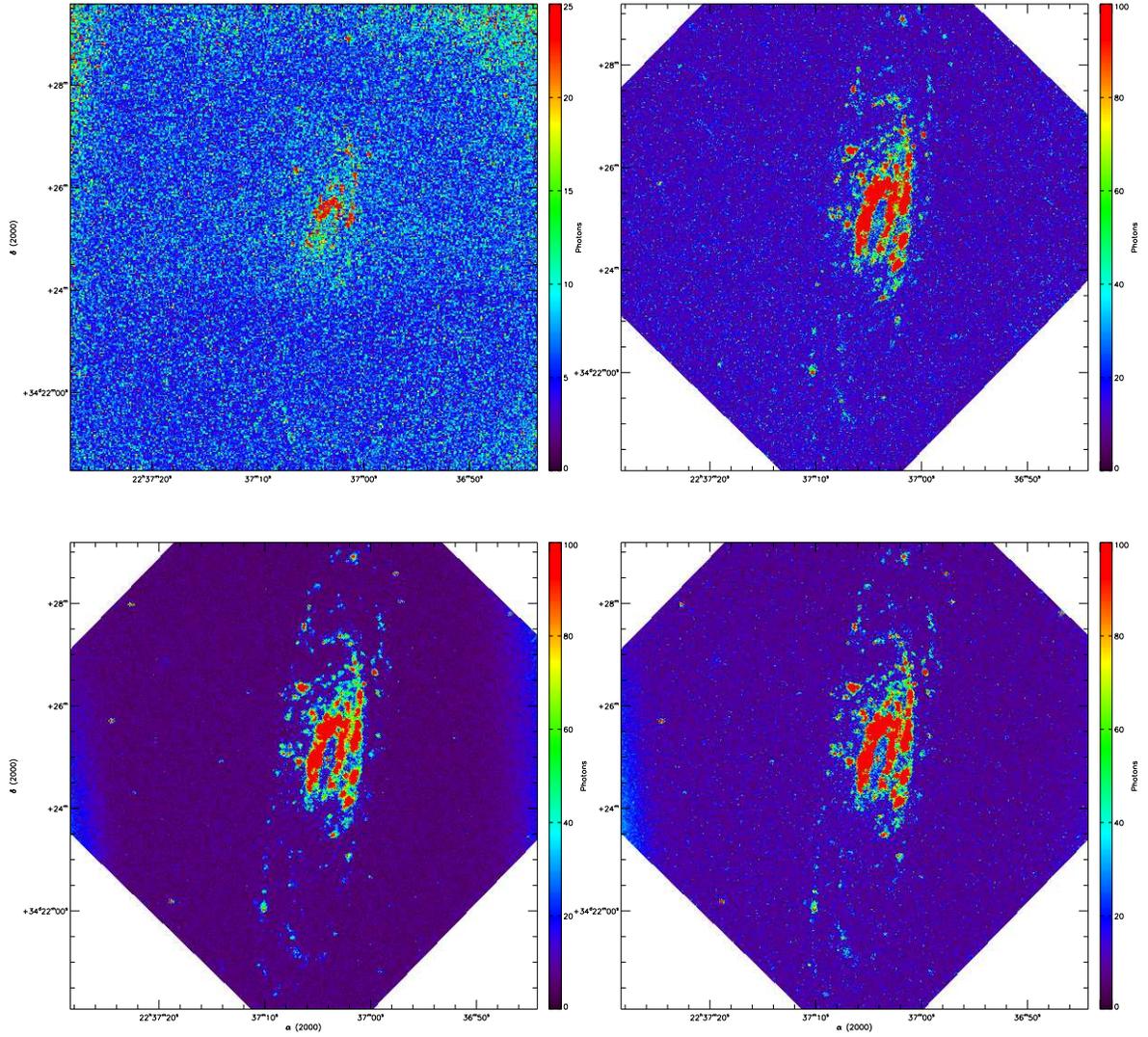}
\caption{Monochromatic intensity maps, after the sky emission removal. \textbf{Top left}: IPCS. \textbf{Top right}: CCCP/CCD97, 5s exposing time per image, AM processing. \textbf{Bottom left}: CCCP/CCD97, 0.5s exposing time per image, PC processing. \textbf{Bottom right}: CCCP/CCD97, 0.5s exposing time per image, AM processing.}
\label{fig::ngc7331Mono}
\end{center}
\end{figure*}

\begin{figure*}[tbp]
\begin{center}
\includegraphics[width=\textwidth]{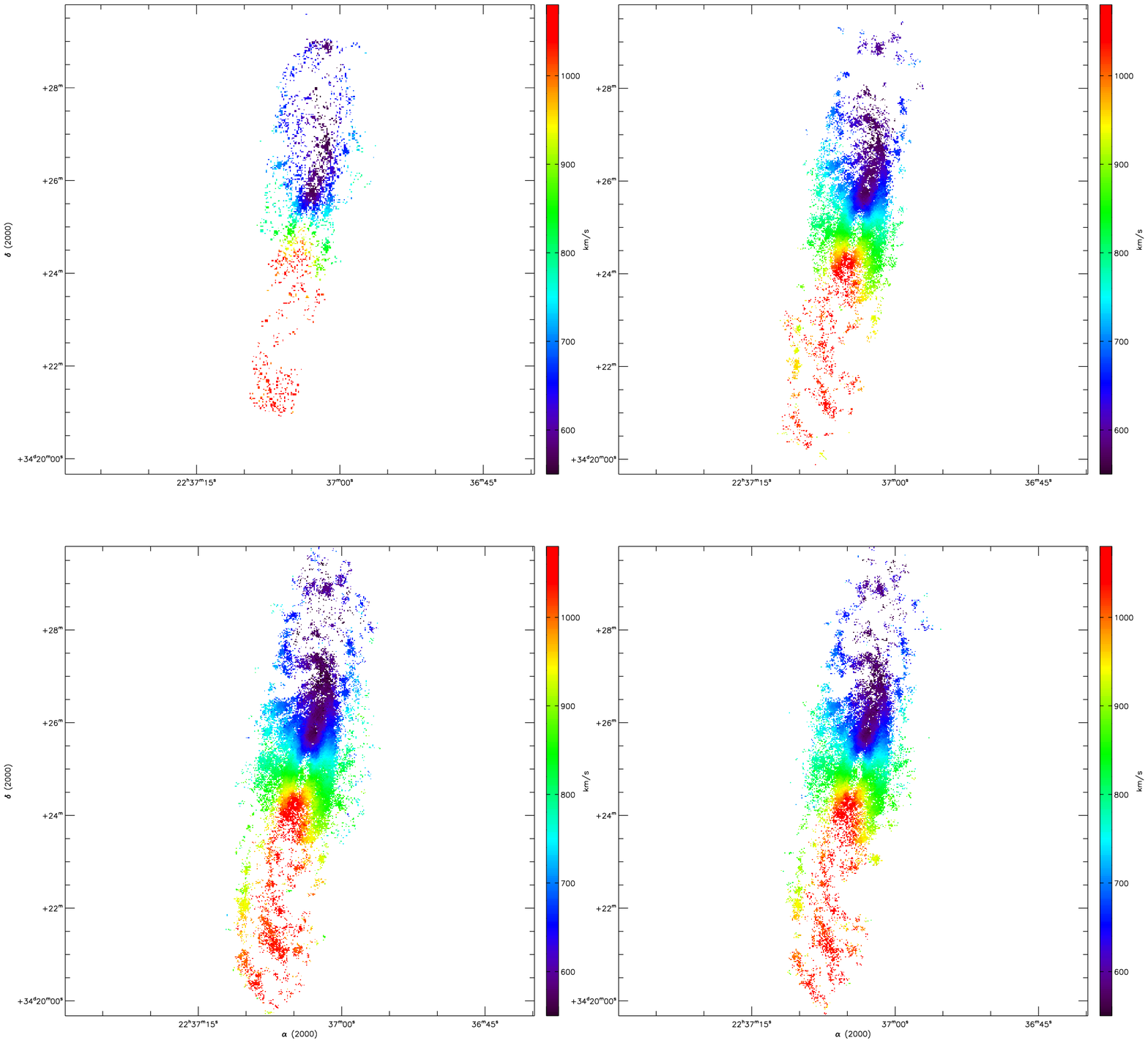}
\caption{RV maps. \textbf{Top left}: IPCS. \textbf{Top right}: CCCP/CCD97, 5s exposing time per image, AM processing. \textbf{Bottom left}: CCCP/CCD97, 0.5s exposing time per image, PC processing. \textbf{Bottom right}: CCCP/CCD97, 0.5s exposing time per image, AM processing.}
\label{fig::ngc7331Rv}
\end{center}
\end{figure*}

On the whole, it is a delicate matter to compare observations made under different environmental and photometric conditions. Nevertheless, the observations should give a first order estimate of how the CCCP/CCD97 compares to the IPCS. The Monochromatic Intensity (MI) maps and the RV maps of all the observations are provided in figures \ref{fig::ngc7331Mono} and \ref{fig::ngc7331Rv}. One striking observation is that, even tough the integration time is comparable between the IPCS and the CCCP/CCD97 observations, the galaxy's approaching side is barely visible on the MI map of the IPCS. This is reflected in the RV map, which shows only a few sparse pixels from this side of the galaxy. By comparing with the CCCP/CCD97 observations, it is questionable whether these pixels are giving an accurate reading of the rotation velocity of this side of the galaxy. Even tough the observation's conditions were different, it is hardly conceivable that the IPCS observation could have gone deeper than the CCCP/CCD97 observation even in perfect conditions. Moreover, the colder temperature at which the IPCS data were acquired should benifit the approaching (red) side as compared to the CCCP/CCD97 observation.

\subsection{Efficiency}
\begin{figure*}[tbp]
\begin{center}
\includegraphics[width=\figurewidth]{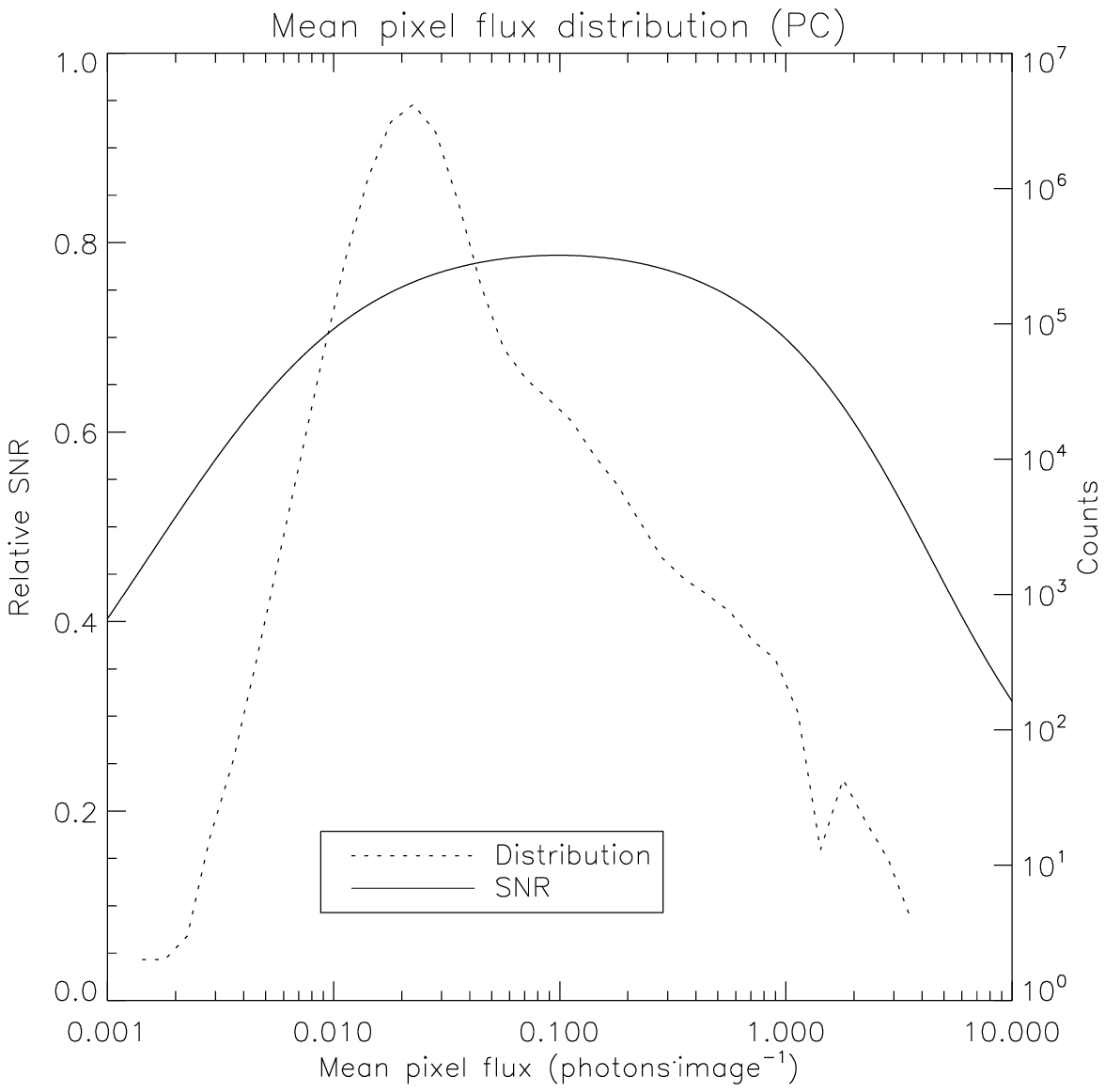}
\includegraphics[width=\figurewidth]{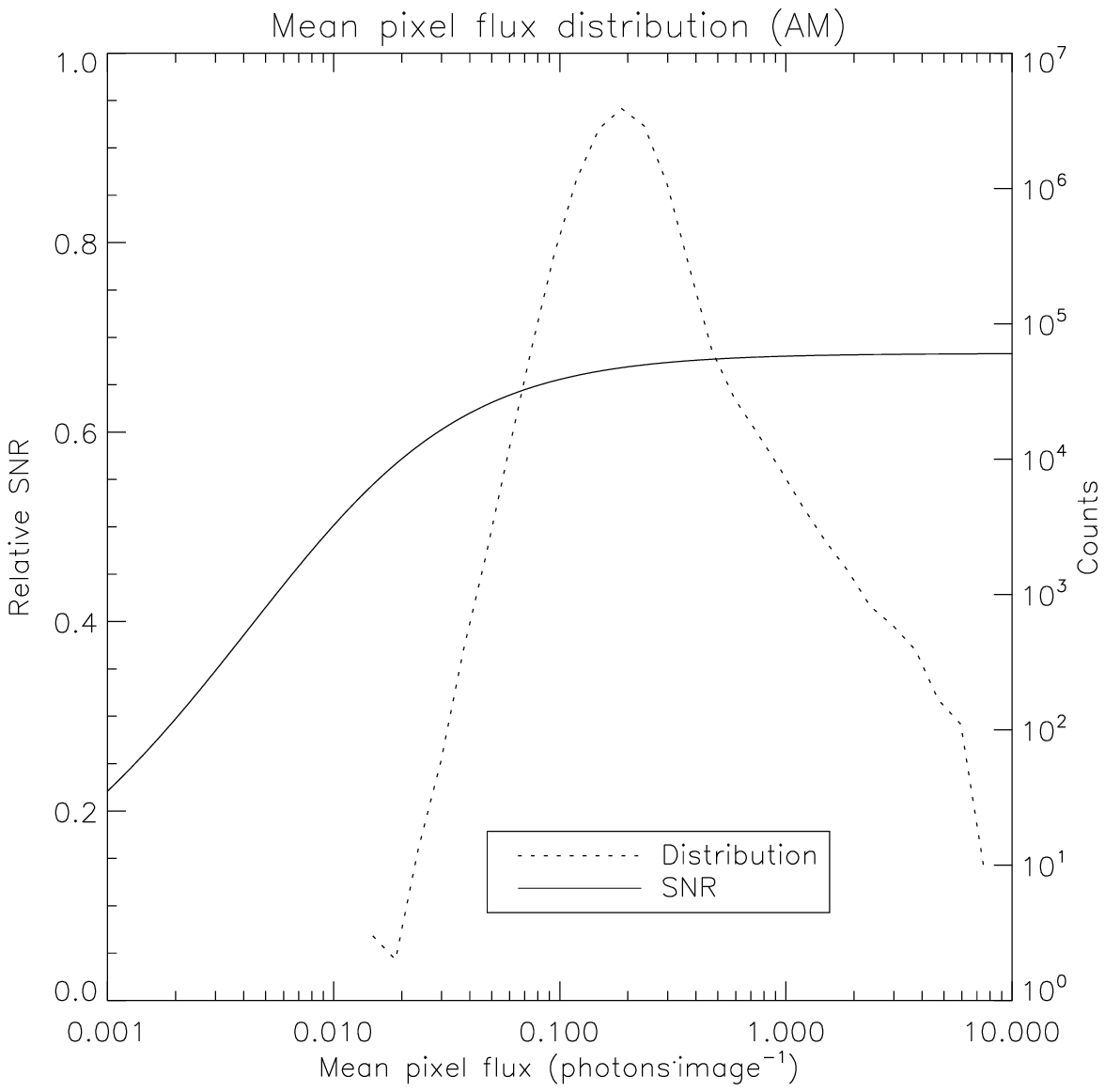}
\includegraphics[width=\figurewidth]{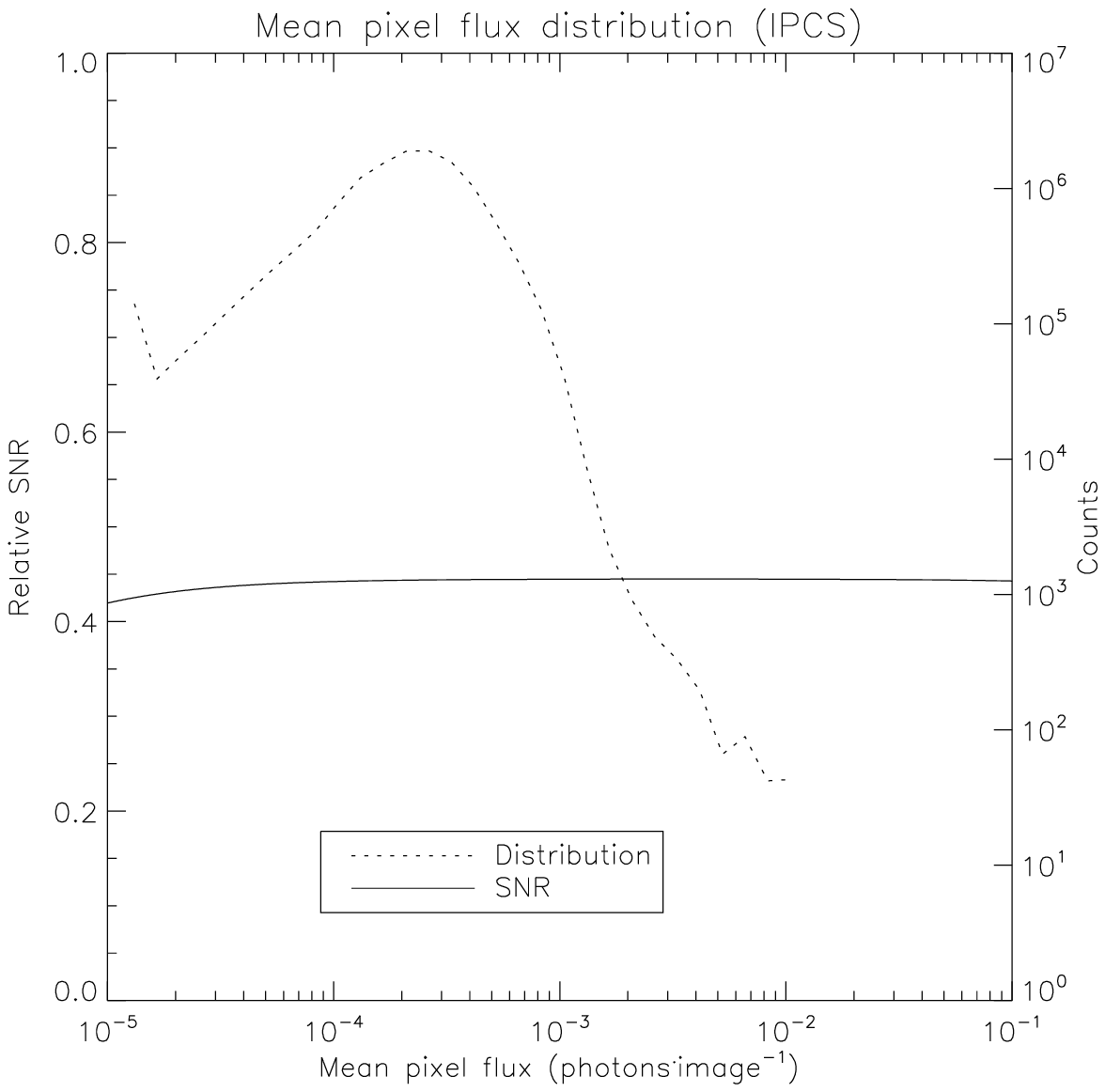}
\caption{Pixel flux distributions of the observation of \mbox{NGC 7331}. The dotted line shows the pixel flux distribution and the plain line shows the relative SNR expected for this flux. The SNR curve includes the QE, CIC, dark and exposing time parameters of table \ref{table::FPobservations}. \textbf{Top left}: PC observation, 0.5s per image. The flux has been corrected for the coincidence losses, which explains the possibility to have more than one photon per image. \textbf{Top right}: AM observation, 5s per image. \textbf{Bottom}: IPCS observation. Given the high frame rate of the IPCS, there is no need to correct the coincidence losses.}
\label{fig::ngc7331Flux}
\end{center}
\end{figure*}

The efficiency at which the data were gathered is outlined by figure \ref{fig::ngc7331Flux}. This figure shows the pixel's mean intensity distribution in the data cubes superimposed on the expected SNR curve for that flux regime. It shows that all the observations were made in roughly the best conditions for the detectors and their modes of operation. The IPCS mean pixel flux is very low: the IPCS is operated at 40 frames s$\mathrm{^{-1}}$ but does not suffer from CIC. Thus, there is no loss at operating that fast.

\subsection{PC versus AM processing}
The observing conditions prevailing when the PC and AM data cubes were gathered were not stable enough to provide a convincing comparison. One could see that the background of the MI map of the PC observation is smoother (figure \ref{fig::ngc7331Mono}). The RV map of the AM observation looks "skinny" as compared to the PC one. On this side, it is better to rely on experimental data taken in lab in order to compare both operating modes with their respective integration time (section \ref{sect::experimentalSNR}).

One comparison that is absolute, however, is the comparison of the PC and AM processing on the data acquired at 0.5 s frame$\mathrm{^{-1}}$ on CCCP/CCD97 (the PC observation). Since there are very few pixels having $\sim$1 photon image$\mathrm{^{-1}}$ (figure \ref{fig::ngc7331Flux}), the effective SNR of the data cube processed in PC should be higher than the one processed in AM. But how does this obvious fact influences the quality of the reduced data of \mbox{NGC 7331}?

\begin{figure*}[tbp]
\begin{center}
\includegraphics[width=\figurewidth]{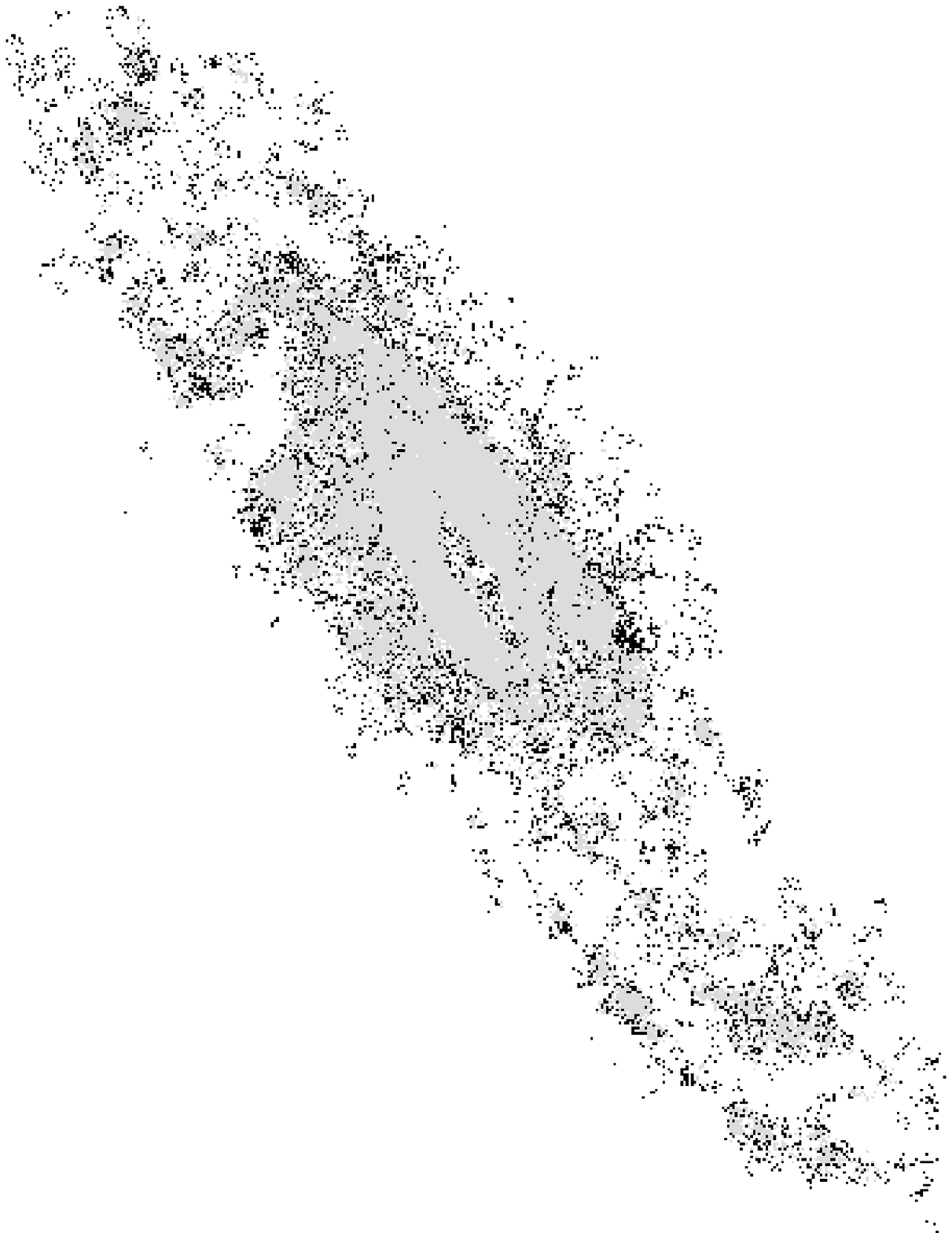}
\includegraphics[width=\figurewidth]{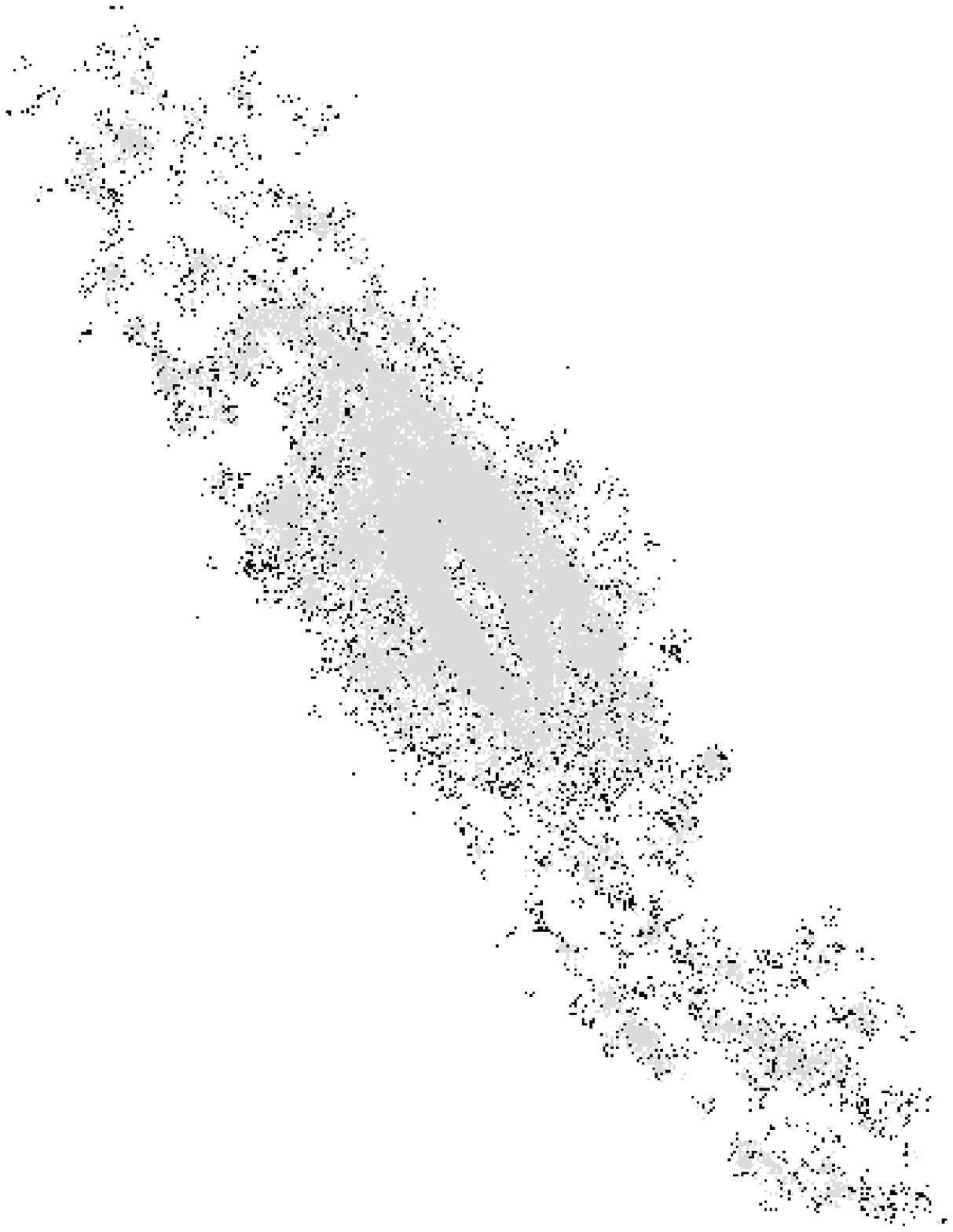}
\caption{Differences between the PC and AM processing of the same observation. The gray pixels represents pixels that are present in the RV maps of both processings. \textbf{Left}: Black pixels represents the pixels that are unique to the PC-processed RV map. \textbf{Right}: Black pixels represents the pixels that are unique to the AM-processed RV map.}
\label{fig::ngc7331NewPixels}
\end{center}
\end{figure*}

The bottom panels of figures \ref{fig::ngc7331Mono} and \ref{fig::ngc7331Rv} shows, respectively, the monochromatic intensity and the RV maps of both the PC and AM processing of the PC observation. First, the monochromatic map of the PC processing shows a smoother and darker background, which means that the sky emission has been better removed. This is very important as it will allow the resolving of the radial velocity of fainter regions of the galaxy. Next, the RV map of the PC processing shows a more extended coverage of the signal. This extended coverage is seen as more pixels near the edge of \hII\, regions and more sparse pixels in diffuse \ha\, regions. Figure \ref{fig::ngc7331NewPixels} enhances these differences by drawing in black the pixels that are exclusive to the PC or AM RV map. There are far more pixels exclusive to the PC RV map than there are for the PC AM map. Thus, as expected, the PC processing yields a better, richer RV map than the AM processing.

\section{Summary and conclusions}
A new camera built using CCCP, a CCD Controller for Counting Photons, and a scientific grade EMCCD CCD97 from e2v technologies was presented. This camera combines the high QE of the CCD with the sub-electron read-out noise of the EMCCD, allowing extreme faint flux imaging at an efficiency that is comparable to a perfect photon counting device. Laboratory experiments showed that the camera behave exactly as predicted by theory, in terms of relative SNR as a function of the flux, down to flux as low as 0.001 photon pixel$\mathrm{^{-1}}$ per image.

It has been shown that the CCCP/CCD97 camera yields very low CIC+dark levels at high EM gain, which represents a significant advance in the sensitivity that an EMCCD camera equipped with a CCD97 can achieve. Moreover, it has been demonstrated that the CTE figure of the EMCCD can be better handled with CCCP. It is beyond the scope of this paper to describe in details the clocking that is applied to the CCD97 to yield the lower CIC and higher CTE. Interested readers are invited to read Daigle et al. (in preparation).

Both PC and AM processing were compared and it has been demonstrated that the PC processing effectively allowed the achievement of a better SNR, for the same data-set. Evidences that the PC processing yields a better SNR than the AM processing, even at 10 times the frame rate, for the same total integration time, was presented. Given that the theory behind the SNR of an EMCCD seems well understood, one could argue that the PC processing should yield a better SNR at low and moderate fluxes, even at a higher frame rate, as compared to AM processing.

The performance of CCCP/CCD97 was compared to that of a GaAs photocathode-based IPCS on an extragalactic target in a shot noise-dominated regime. The achieved SNR of the CCCP/CCD97 observations is superior to the one achieved with the IPCS. Though it is delicate to compare two independent observations taken at different photometric conditions, the results presented could hardly be due solely to the photometric conditions. The CCCP/CCD97 camera must be, at some level, more sensitive than the IPCS, for the flux regime at which the data were acquired.

A new camera that will use the second version of the CCCP controller is under construction at the LAE in Montr\'eal. It will use a larger 1600 $\times$ 1600, non frame transfer, EMCCD from e2v Technologies. This camera will be used as the high resolution camera at the focal plane of the 3D-NTT instrument \citep{2008SPIE.7014E.170M} that is expected to make first light at the end of 2009.

\acknowledgments

O. Daigle is grateful to the NSERC for funding this study though its Ph. D. thesis. We would like to thank the staff at the Observatoire du mont M\'egantic for their helpful support, and the anonymous referee for valuable comments.



{\it Facilities:} \facility{Observatoire du mont M\'egantic (OMM)}, \facility{Laboratoire d'Astrophysique Exp\'erimentale (LAE)}.





\clearpage



\end{document}